\documentclass[aps,prx,twocolumn,showpacs,superscriptaddress,longbibliography]{revtex4-2}

\usepackage[utf8]{inputenc}
\usepackage{amsmath}
\usepackage{graphicx}
\usepackage{lipsum}
\usepackage{braket}
\usepackage{xcolor}
\usepackage{xstring}
\usepackage{tikz}
\usepackage{bbm}
\usepackage{babel}

\usepackage[bookmarksnumbered,pdfpagelabels=true,plainpages=false,colorlinks=true,linkcolor=blue,citecolor=blue,urlcolor=blue]{hyperref}

\usepackage{txfonts}

\newcommand{\GF}{\hat{\mathcal G}}
\newcommand{\SE}{\hat\Sigma}
\newcommand{\SF}{\mathcal A}
\DeclareMathOperator{\re}{\mathrm{Re}}
\DeclareMathOperator{\im}{\mathrm{Im}}
\DeclareMathOperator{\tr}{\mathrm{Tr}}
\DeclareMathOperator{\diag}{\mathrm{diag}}
\newcommand{\br}[1]{\left(#1\right)}

\newcommand{\sumBZ}[1]{\sum_{#1\in\mathrm{BZ}}}
\newcommand{\nn}[1]{\left<#1\right>}
\newcommand{\nnn}[1]{\left<\left<#1\right>\right>}
\newcommand{\hc}{h.c.}

\newcommand{\captionSpectraZigzag}{
    Single-magnon spectra and two-magnon density of states for the strip with zig-zag edge termination ($N_y = 30$, $D_\parallel = 0.1 JS$, numerical line broadening $0.02JS$). 
    (a) Single-magnon spectrum without magnon-magnon interactions ($D_\perp = 0J\sqrt S$); inset (d) shows a zoom into the sharp chiral edge mode, visible inside the topological band gap around the center of the Brillouin zone.
    (b) Single-magnon spectrum with strong magnon-magnon interactions ($D_\perp = 0.4J\sqrt S$).
    The bulk modes of the upper band acquire a strong lifetime broadening and inset (e) shows that the chiral edge modes disappear completely.
    (c) Two-magnon density of states describing the available decay phase space for a magnon with initial momentum $k_x$ and energy $\omega$.
    The chiral edge modes from panel (a) lie in a region of large two-magnon density of states, which enhances their lifetime broadening.
    Panels (a)-(e) additionally show the edge mode (band edge) dispersions obtained from $\hat H_2$ (LSWT) as dashed red (green) lines.
    (f) Spectral functions at the reference momentum $k_x^*$ [dashed white line in (d,e)] versus frequency inside the band gap for different magnon-magnon interaction strengths $D_\perp$.
    For visual guidance, the peaks for $D_\perp = 0.0, 0.2 J \sqrt S$ are decorated with Lorentzian fits.
}
\newcommand{\captionSpectraDangling}{
    Same as Fig.~\ref{fig:spectra_zigzag} but for dangling edge termination.
}
\newcommand{\captionDecayRates}{
    Edge mode decay rates for different interaction strengths $D_\perp$ along selected momentum cuts through the Brillouin zone.
    Decay rates obtained from the on-shell approximation are depicted with solid lines, while dotted lines correspond to decay rates obtained self-consistently from the imaginary parts of the Green function poles.
    (a, b) non-interacting band structures obtained from $\hat H_2$ for dangling and zig-zag edge terminations, respectively.
    The edge modes whose decay rates are shown in panels (c) and (d) are highlighted in red along the respective momentum cuts.
    (c) Decay rates of the higher-energy edge mode for dangling edge termination between the $\Gamma$- and the $K$-point.
    The reference momentum $k_x^*$ from Figs.~\ref{fig:spectra_dangling}(d,e) is also indicated.
    (d) Decay rates of the left edge mode for zig-zag edge termination in the vicinity of the $K$-point.
    We also show the reference momentum $k_x^*$ from Figs.~\ref{fig:spectra_zigzag}(d,e).
}
\newcommand{\captionSpectraRealSpace}{
    With dangling edge termination, the higher-energy edge mode delocalizes due to hybridization with bulk modes.
    (a) $y$-resolved spectral functions on the $B$-sublattice evaluated at the reference momentum $k_x^*$ from Fig.~\ref{fig:spectra_dangling} versus frequency within the band gap for different $y$-coordinates at interaction strength $D_\perp = 0.4 J \sqrt S$.
    The visible edge mode peaks are equipped with Lorentzian fits.
    The cylinder on the right is a sketch of the system for periodic boundary conditions in $x$-direction and open boundary conditions in $y$-direction; the chosen $y$-coordinates are highlighted by rings of the corresponding colors.
    Note that the plot does not show any signature of the lower-energy edge mode because it is predominantly localized on the $A$-sublattice at the lower end of the cylinder.
    Inset: spectral function from Fig.~\ref{fig:spectra_dangling} with a gray bar at $k_x^*$ indicating the frequency window on the horizontal axis of the main plot.
    (b) Quasiparticle weights extracted from Lorentzians fitted to the edge mode peaks versus $y$ for different interaction strengths.
    At finite interaction strengths $D_\perp > 0 J\sqrt S$, the exponential decay at the edge levels off at a roughly constant value inside the bulk, approximated by the delocalization parameter $\Delta$.
    Inset: delocalization parameter versus interaction strength.
}

\newcommand{\captionDecayRatesZigzagBField}{
    Restoration of a sharp edge mode peak through a magnetic field for the zig-zag edge termination.
    (a) Spectral functions evaluated at the reference momentum $k_x^*$ from Fig.~\ref{fig:spectra_zigzag} versus frequency within the band gap for different magnetic fields $B$.
    The magnetic field shifts the band gap to higher frequencies while gradually restoring the spectral peak of the left edge mode.
    Lorentzians are fitted to the visible edge mode peaks for visual guidance.
    Inset: non-interacting band structure obtained from $\hat H_2$ with a black bar indicating indicating the frequency window on the horizontal axis of the main plot.
    Notice that the inset shows $\omega - B$ on the vertical axis.
    (b) Decay rates of the left edge mode at momentum $k_x^*$ versus magnetic field for different interaction strengths.
    Solid lines are the on-shell values, dotted lines are self-consistently extracted from the Green function poles.
}

\newcommand{\captionSpectraZigzagBField}{
    Single-magnon spectral functions and two-magnon densities of states for zig-zag edge termination at representative values of the magnetic field $B$.
    Non-interacting band edges and edge mode dispersions are overlaid like in Figs.~\ref{fig:spectra_zigzag} and \ref{fig:spectra_dangling}.
    At finite magnetic field, the single-magnon spectra are shifted upwards by $B$ while the two-magnon densities of states are shifted upwards by $2B$ compared to the $B=0$ case.
    At about $B=1.0JS$, the areas of high two-magnon density of states start leaving the band gap, restoring sharp and long-lived edge modes.
}

\newcommand{\captionEdgeBulkOverlapVsSlabSize}{
    Edge-bulk hybridization strength $w$ versus number of unit cells in $y$-direction for interaction strength $D_\perp = 0.4J\sqrt S$ at zero magnetic field.
    The data points approach a constant $\approx 0.016$ in the thermodynamic limit, signifying a stable edge-bulk hybrid state extending over the whole system.
}

\newcommand{\figTwoColumn}[2]{
    \begin{figure*}[t!]
        \includegraphics[width=\textwidth]{figures/#1}
        \caption{#2}
        \label{fig:#1}
    \end{figure*}
}

\newcommand{\fig}[2]{
    \begin{figure}[t!]
        \includegraphics[width=0.5\textwidth]{figures/#1}
        \caption{#2}
        \label{fig:#1}
    \end{figure}
}

\begin{document}

\title{
Breakdown of Chiral Edge Modes in Topological Magnon Insulators
}

\author{Jonas Habel}
\affiliation{Technical University of Munich, TUM School of Natural Sciences, Physics Department, 85748 Garching, Germany}
\affiliation{Munich Center for Quantum Science and Technology (MCQST), Schellingstr. 4, 80799 M{\"u}nchen, Germany}
\author{Alexander Mook}
\affiliation{Institute of Physics, Johannes Gutenberg-University Mainz, Staudingerweg 7, Mainz 55128, Germany}
\affiliation{Technical University of Munich, TUM School of Natural Sciences, Physics Department, 85748 Garching, Germany}
\author{Josef Willsher}
\affiliation{Technical University of Munich, TUM School of Natural Sciences, Physics Department, 85748 Garching, Germany}
\affiliation{Munich Center for Quantum Science and Technology (MCQST), Schellingstr. 4, 80799 M{\"u}nchen, Germany}
\author{Johannes Knolle}
\affiliation{Technical University of Munich, TUM School of Natural Sciences, Physics Department, 85748 Garching, Germany}
\affiliation{Munich Center for Quantum Science and Technology (MCQST), Schellingstr. 4, 80799 M{\"u}nchen, Germany}
\affiliation{Blackett Laboratory, Imperial College London, London SW7 2AZ, United Kingdom}

\date{August 6, 2023}
\begin{abstract}
Topological magnon insulators (TMI) are ordered magnets supporting chiral edge magnon excitations. These edge states are envisioned to serve as topologically protected information channels in low-loss magnonic devices. The standard description of TMI is based on  linear spin-wave theory (LSWT), which approximates magnons as free non-interacting particles. However, magnon excitations of TMI are genuinely interacting even at zero temperature, calling into question descriptions based on LSWT alone. Here we perform a detailed \emph{non-linear} spin-wave analysis to investigate the stability of chiral edge magnons. We identify three general breakdown mechanisms: (1) The edge magnon couples to itself, generating a finite lifetime that can be large enough to lead to a \emph{spectral annihilation} of the chiral state; (2) The edge magnon hybridizes with the extended bulk magnons and, as a consequence, \emph{delocalizes} away from the edge; (3) Due to a bulk-magnon mediated \emph{edge-to-edge coupling}, the chiral magnons at opposite edges hybridize. We argue that, in general, these breakdown mechanisms may invalidate predictions based on LSWT and violate the notion of topological protection. We discuss strategies how the breakdown of chiral edge magnons can be avoided, e.g.~via the application of large magnetic fields. Our results highlight a challenge for the realization of chiral edge states in TMI and in other bosonic topological systems without particle number conservation.
\end{abstract}

\maketitle

\section{Introduction}
\label{sec:intro}

Magnon spintronics has emerged as a promising candidate to meet today's grand challenge of realizing computational technologies with a minimal environmental footprint \cite{chumak2015magnon}. Encoding information in the excitations of magnetically ordered insulators, called magnons, bypasses the Joule heating associated with moving electrons, thus facilitating low-power computing. To increase the lifetime of magnonic signals, it has recently been proposed to exploit topological protection of magnons \cite{katsura2010, onose_observation_2010,matsumoto_rotational_2011,matsumoto_theoretical_2011, Shindou2013, Hoogdalem2013, zhang_topological_2013,mook_magnon_2014,mook_edge_2014,chisnell_topological_2015,hirschberger_thermal_2015, Nakata2017QHE, zhang_interplay_2021}. Topological magnon insulators (TMI) have been suggested to support chiral edge magnons that are immune to backscattering, similar to the chiral electronic edge states in Chern insulators. The chiral magnon edge states are the key ingredient for novel spintronic devices such as spin-wave beam splitters, Fabry--Perot interferometers, and waveguides \cite{Shindou2013,Mook2015a, Mook2015b, Wang2018, Wang2021}. The proposal of TMI in magnetic materials gave rise to the flourishing field of magnon topology recently reviewed in Refs.~\onlinecite{Malki2020, Kondo2020review, Li2021, Bonbien2021, McClartyReview2022}.

\begin{figure}
    \centering
    \includegraphics[width=1\columnwidth]{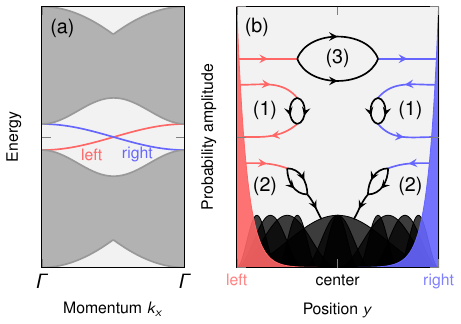}
    \caption{
    Conceptual illustration of how the protection of chiral edge states in TMI breaks down.
    Consider a TMI
    with periodic (open) boundary conditions in the $x$ ($y$) direction. (a) At the level of LSWT, the single-magnon spectrum exhibits a gap between the bulk bands, which are denoted by the gray areas. This gap supports chiral edge states, indicated by red and blue colors, which propagate in opposite directions at the opposite edges (denoted left and right). (b) The  probability amplitudes of the edge states are exponentially localized to their respective edges. However, at the level of nonlinear spin-wave theory, there are several many-body interaction processes (sketched as Feynman diagrams) that can couple the edge magnons to any other magnons via virtual transitions through the two-magnon continuum. There are three main processes that (1) couple an edge magnon to itself, (2) couple edge magnons to the bulk magnons (denoted by the black extended probability amplitudes), and (3) couple the opposite edge magnons. 
    Process (1) leads to a finite lifetime of edge magnons, which can be large enough to annihilate the spectral weight of the edge state. We call this process \emph{spectral annihilation}. Process (2) leads to a hybridization of edge magnons with bulk magnons. As a result, the edge magnons \emph{delocalize} into the bulk.
    Process (3) causes a hybridization of states on opposite edges, violating their topological protection via \emph{edge-to-edge coupling}. 
    Taken together, the three processes can violate all distinguishing features of topologically protected chiral edge states, invalidating the conclusions of LSWT.}
    \label{fig:teaser}
\end{figure}

Among the first examples of TMI were the honeycomb ferromagnets with next-nearest-neighbor 
Dzyaloshinskii--Moriya interactions (DMI) \cite{owerre_first_2016, Kim2016}. 
Several materials, including $\mathrm{CrI_3}$, CrGeTe$_3$, and CrSiTe$_3$, are thought to be described by variations of this model, as supported by evidence from several inelastic neutron scattering and thermal Hall experiments \cite{chen_topological_2018,zhu_topological_2021,zhang_anomalous_2021}.
In an approximate semiclassical description, the linear spin wave theory (LSWT), the honeycomb ferromagnets host topological magnon bands whose properties, including Berry curvature and Chern number, closely resemble the electronic bands in the Haldane model \cite{owerre_first_2016,Kim2016,owerre_topological_2016}.
At the LSWT level, this analogy between magnonic and electronic band topology suggests that there are protected chiral magnon edge modes due to the bulk-boundary correspondence, as depicted in Fig.~\ref{fig:teaser}(a).
Importantly, it is often implicitly assumed that these edge modes carry over to the full spin model, based on the experience that the LSWT is a reasonably good approximation for many materials, especially ferromagnets \cite{dyson_thermodynamic_1956,auerbach1998interacting}.
As a result, the field of magnon topology relies heavily on the LSWT, which has become the standard for classifying and searching for new topological magnon materials \cite{Corticelli2022, Corticelli2022b, Karaki2023}, for interpreting inelastic neutron scattering data in the context of topology \cite{chisnell_topological_2015, chen_topological_2018,  Elliot2021, zhu_topological_2021, zhang_anomalous_2021, Scheie2022, Nikitin2022}, and for understanding thermal Hall effect data in terms of a magnon Berry curvature \cite{onose_observation_2010, katsura2010, matsumoto_rotational_2011, matsumoto_theoretical_2011, Ideue2012, hirschberger_thermal_2015, zhang_anomalous_2021, Akazawa2022, Neumann2022, czajka2023planar}.

However, explicit experimental evidence for topological chiral magnon edge modes in quantum magnets is lacking. Conventional probes, such as inelastic neutron scattering, measure only the bulk magnon spectrum.
Also, measurements of transverse transport properties cannot reliably detect magnon topology as is possible for electronic topological insulators, where quantized topological markers are accessible.
This difference is due to the Bose statistics of magnons: their contribution to the thermal Hall effect is neither quantized nor does it vanish in topologically trivial magnonic systems  \cite{katsura2010,matsumoto_rotational_2011}.
It has been proposed to detect chiral magnon edge modes by amplification of their occupation via driving with electromagnetic fields \cite{malz_topological_2019}.
Similarly, other theoretical proposals for the direct detection of chiral magnon edge states have been put forward \cite{Perreault2016, Rustagi2020, Feldmeier2020, Guemard2022, Hetenyi2022, Mitra2023, Gunnink2023}, but have not yet been experimentally implemented.

The fact that, despite much effort, topological edge magnons have not been observed motivates a detailed microscopic analysis of their stability.
Indeed, we find that a closer theoretical investigation reveals several potential breakdown mechanisms of chiral edge magnons.
First of all, the LSWT used to describe the edge modes is an approximation and its validity must be carefully checked.
In the case of antiferromagnets, the non-interacting LWST approximation has long been understood to have limitations in predicting magnon spectra \cite{elliott_effects_1969,auerbach1998interacting}.
Moreover, recent theoretical work suggests that magnon-magnon interactions beyond the LSWT can significantly alter the bulk excitation spectrum in systems with anisotropic interactions such as DMI or non-collinear spin textures \cite{chernyshev_spin_2009,maksimov_rethinking_2020,zhitomirsky_colloquium_2013}. 
In the context of TMI, such magnon-magnon interactions have been shown to induce strong decay of the bulk magnons in kagome magnets with DMI, potentially undermining the bulk topology and jeopardizing the integrity of chiral edge states \cite{chernyshev_damped_2016}. For field polarized honeycomb  Kitaev magnets, there is numerical and analytical evidence of TMI regimes  with stable chiral egde modes \cite{mcclarty_topological_2018}. 
In this work, we address in detail the general stability of chiral edge modes beyond LSWT.

The key point is that the analogy between magnonic and electronic Chern insulators generally holds only up to LSWT.
Beyond this approximation, fundamental differences between magnons and electrons come into play, which make magnonic chiral edge modes susceptible to decay due to many-body interactions always present in spin systems.
To have such decays, the scattering between states must be both (1) \emph{kinematically allowed} and (2) permitted by \emph{non-zero matrix elements} \cite{zhitomirsky_colloquium_2013}.
Both ingredients are missing for chiral edge modes in electronic topological insulators \cite{hasan_colloquium_2010}, but can be \textit{a priori} present in TMI.

First, regarding the kinematic condition~(1), we recall that particles of electronic topological insulators are fermions with a chemical potential in the band gap.
Therefore, edge modes are low-energy degrees of freedom protected from decay because the density of available final states vanishes at low energies and temperatures.
In contrast, magnons are bosonic, having zero chemical potential.
The chiral edge modes are, thus, high-energy degrees of freedom connecting bulk bands, which have significant energetic overlap with the two- or multi-magnon continuum in many materials.
As a result, the decay of the magnon chiral edge modes is often kinematically allowed.

Second, regarding the quantum mechanical condition~(2), note that with number-conserving magnon-magnon interactions in the Hamiltonian --- as in a simple Heisenberg ferromagnet --- decay can only occur if mediated by thermally excited magnons \cite{dyson_general_1956,dyson_thermodynamic_1956, Pershoguba2018, Lu2021, Nikitin2022, Sun2023}.
However, there is no general particle number conservation of magnons --- for example, DMI interactions are enough to violate it \cite{chernyshev_damped_2016, mcclarty_non-hermitian_2019, mook_interaction-stabilized_2021, Gohlke2022, Mook2023multi} and potentially induce strong \emph{spontaneous} magnon decay even at zero temperature \cite{zhitomirsky_colloquium_2013}. For general models of TMIs, the spin-orbit like spin exchange opening the topological gaps simultaneously facilitates magnon decay.

The expectation that chiral edge magnons enjoy a degree of ``topological protection'' is based on the following three features found at the level of LSWT: (i) they have an infinite lifetime, (ii) they are exponentially localized to their respective edges, as can be seen from their probability amplitudes, and (iii) they do not hybridize with the edge state at the opposite edge in the thermodynamic limit -- more precisely, their hybridization gap in finite systems decays exponentially in the system size over their localization length.
Here, concentrating on the paradigmatic TMI of the honeycomb-ferromagnet with DMI, we show how features (i) and (ii) may disappear and feature (iii) may be severely undermined in the presence of ubiquitous multi-magnon interactions.
We argue that there are in principle \emph{three breakdown mechanisms} of chiral edge magnons, as depicted schematically in Fig.~\ref{fig:teaser}(b).
\begin{enumerate}
    \item \emph{Spectral annihilation:} Chiral edge magnons undergo a strong spontaneous decay that leads to a spectral lifetime broadening. This effect can be so strong that, in some situations, they disappear completely.
    \item \emph{Delocalization:} Chiral edge magnons hybridize with the two-magnon continuum, which is made up of bulk modes at lower energies, as well as with single-magnon bulk modes. As a result, their wavefunction gains significant support deep in the bulk, leading to a delocalization from the edge.
    \item \emph{Edge-to-edge coupling}: States on opposite edges may hybridize via coupling to extended two-magnon bulk modes, violating their topological protection.
    In particular, the decay of the hybridization gap, albeit still exponential in system size, can be orders of magnitude slower than that predicted in the LSWT.
\end{enumerate}

Our main finding is that predictions of topological edge magnons based on LSWT are at least imprecise, and under many circumstances any notion of ``topological protection" can be qualitatively wrong. On a more positive note, we discuss under which conditions the results of LSWT can be stabilized, and confirm the limit of large magnetic fields \cite{mcclarty_topological_2018} as the most promising regime for stable topological chiral edge magnons.

The rest of the paper is organized as follows. In Sec.~\ref{sec:model}, we present the topological honeycomb ferromagnet with DMI, discuss the LSWT and magnon-magnon interactions for this model on a bounded strip geometry, and apply diagrammatic perturbation theory to the topological magnon edge modes.
Section~\ref{sec:renormalized_spectra} presents zero-temperature single-magnon spectral functions in the strip geometry with open boundary conditions, showing the strong many-body renormalization of the edge modes.
To demonstrate the non-universality of these effects, we compute the spectral functions for both ``zig-zag" and ``dangling" edge terminations.
Section~\ref{sec:spontaneous_decay_rates} is devoted to quantifying the decay rates of the edge modes using both a self-consistent solution of the pole condition, and the widely-used on-shell approximation. The performance of these two methods is consequently compared.
Section~\ref{subsec:edge_bulk_hybridization} discusses the hybridization of edge states with single-magnon and two-magnon bulk modes.
We show how the edge state of the LSWT becomes an edge resonance with finite support in the bulk due to magnon-magnon interactions.
In Sec.~\ref{subsec:edge_edge_hybridization}, we explore the hybridization of edge states with each other by coupling to the delocalized two-magnon continuum.
In Sec.~\ref{sec:magnetic_fields}, we show how to restore stable and well-defined edge modes by applying a magnetic field.
Finally, in Sec.~\ref{sec:relevance}, we discuss our results and their general implications in a wide variety of ordered magnets and other bosonic systems.
Appendixes~\ref{sec:edge_terminations}-\ref{app:deloc_param_perturbative} provide detailed technical information.

\section{Chiral Edge Modes in a Honeycomb Topological Magnon Insulator}

\label{sec:model}

\begin{figure}[t]
    \centering
    \includegraphics[width=0.5\textwidth]{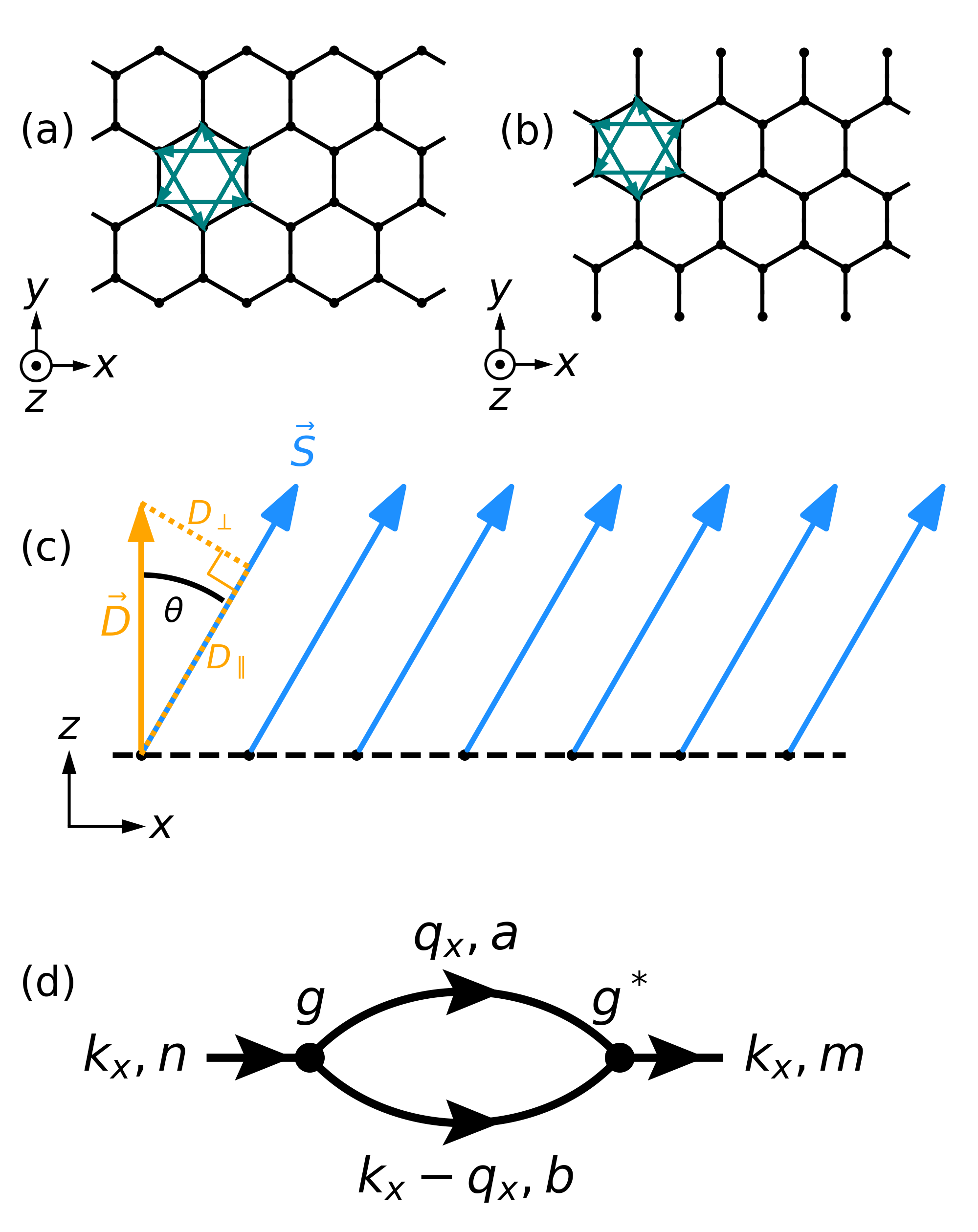}
    \caption{Sketch of the honeycomb lattice on the strip geometry with (a) zig-zag and (b) dangling edge termination in the $y$-direction and periodic boundary conditions in the $x$-direction. Next-nearest-neighbor bonds and their orientations are indicated with teal arrows. In our numerical analysis, the strip has $2N_y = 60$ sites in the $y$-direction unless indicated otherwise. (c) Side view of the honeycomb lattice with the classical ferromagnetic ground state indicated with blue arrows. The magnetization is tilted with respect to the out-of-plane DMI vector $\vec D$ by an angle $\theta$.
    (d) The only non-zero self-energy diagram at one-loop order and zero temperature. $k_x$ ($q_x$) is the external (internal) momentum, and $n,m$ ($a,b$) are the external (internal) band indices. The three-magnon vertex factor $g$ is defined in Eq.~\eqref{eq:cubic_vertex_factors}.}
    \label{fig:lattice_details}
\end{figure}

For concreteness, we consider a spin model on the honeycomb lattice with nearest-neighbor ferromagnetic Heisenberg couplings and next-nearest-neighbor DMI given by
\begin{align}
\label{eq:spin_hamiltonian}
    \hat H = -J \sum_{\braket{ij}} \vec S_i \cdot \vec S_j + \sum_{\braket{\braket{ij}}} \vec D_{ij} \cdot \br{\vec S_i \times \vec S_j},
\end{align}
where the spin operators $\vec S_i$ describe spin-$S$ degrees of freedom, and $J>0$ to stabilize the ferromagnetic order.
Since the honeycomb plane is a mirror plane of the system, the DMI vector must point out of the plane, i.e.\ $\vec D_{ij} = D_{ij} \vec e_z$ \cite{gao_thermal_2019,moriya_anisotropic_1960}, where $\vec e_z$ is a unit vector in the $z$ direction. The sign of the DMI constant is linked to the bond direction. We set $D_{ij} = \lambda_{ij} D$ with $\lambda_{ij} = \pm 1$, where the plus (minus) sign applies if the bond from site $i$ to site $j$ is in the clockwise (counter-clockwise) direction for a given hexagon.
Treated in LSWT, this Hamiltonian exhibits a magnon band structure with gapped Dirac cones due to the DMI, resulting in magnon bands with non-trivial Chern numbers and chiral magnon modes at the edges of the system \cite{owerre_first_2016, Kim2016}.

To study these chiral edge modes, we consider the model on a quasi-1D strip geometry with periodic boundary conditions in the $x$-direction and open boundary conditions in the $y$-direction.
We denote the number of two-site honeycomb unit cells in the $y$-direction as $N_y$.
This geometry has two edges parallel to the $x$-axis [see Fig.~\ref{fig:lattice_details}(a,b)].
The termination of the honeycomb lattice at these edges is not unique and affects the LSWT edge mode dispersion.
To investigate how this in turn influences the many-body magnon physics, we study two different edge terminations: ``zig-zag" edges and ``dangling" edges [see Figs.~\ref{fig:lattice_details}(a) and \ref{fig:lattice_details}(b), respectively]. Additional information about the edge terminations is provided in App.~\ref{sec:edge_terminations}.

The ground state of Hamiltonian \eqref{eq:spin_hamiltonian} is not obvious because there is a competition between the Heisenberg interaction, which aligns the spins in parallel, and the DMI, which favors in-plane spin canting.
However, a numerical analysis shows that, as long as $D \lesssim 0.7JS$, the classical ground state on the strip is a fully polarized ferromagnet and the DMI is completely frustrated [see Fig.~\ref{fig:lattice_details}(c)] \cite{luttinger_note_1951}.
This result is independent of the edge termination (see App.~\ref{app:classical_GS_strip} for details).
Similar to the case of Ref.~\onlinecite{chernyshev_damped_2016}, the ground state polarization direction is arbitrary and spontaneously breaks the Hamiltonian's $\mathrm{SO}(2)$ spin rotation symmetry about the $z$-axis.
For later reference, we define the angle $\theta$ between the magnetization and the out-of-plane $z$-direction, and decompose the DMI vector $\vec D$ into its components parallel, $D_\parallel = D\cos\theta$, and perpendicular, $D_\perp = D\sin\theta$, to the magnetization, as depicted in Fig.~\ref{fig:lattice_details}(c).

\subsection{Spin-Wave Theory}
\label{subsec:spin_wave_theory}

The standard way to describe magnon excitations above the ordered ground state is to map the spin operators to bosonic creation and annihilation operators by a Holstein--Primakoff transformation \cite{holstein_field_1940}:
\begin{align}
    S_i^\parallel &= S - a_i^\dagger a_i, \\
    S_i^+ &= \sqrt{2S - a_i^\dagger a_i}\, a_i.
\end{align}
Here, $S^\parallel_i$ denotes the component of the spin at site $i$ that is parallel to the ferromagnetic polarization direction, $S^+_i$ ($S^-_i = (S^+_i)^\dagger$) is the spin raising (lowering) operator at site $i$ with respect to the polarization direction, and $a_i$ ($a^\dagger_i$) annihilates (creates) a magnon at site $i$.
Plugging this transformation into the Hamiltonian \eqref{eq:spin_hamiltonian} and performing a large-$S$ expansion of the square root terms, the Hamiltonian becomes a power series in the small parameter $1/S$:
\begin{align}
\label{eq:large_S_expansion}
    \hat H = E_\mathrm{GS} + \hat H_2 + \hat H_3 + \hat H_4 + \dots,
\end{align}
where $E_\mathrm{GS}$ is the classical ground state energy and $\hat H_n$ is of order $\mathcal O(S^{2-n/2})$, containing only terms with $n$ bosonic operators.
The quadratic component is the LSWT Hamiltonian
\begin{align}
    \hat H_2 = -JS \sum_{\nn{ij}} \br{a_i^\dagger a_j + \hc} + D_\parallel S \sum_{\nnn{ij}} \br{i a_i^\dagger a_j + \hc} \nonumber \\
    + JS \sum_i z_i a_i^\dagger a_i,
\end{align}
which describes non-interacting magnons. It is a bosonic analog of the Haldane model \cite{haldane_model_1988} with a staggered $\pi/2$-flux through every plaquette and a topological band gap of $6\sqrt 3 D_\parallel S$.
Furthermore, $z_i$ is the number of nearest neighbors of site $i$, i.e., $z_i = 3$ in the bulk, and $z_i \le 3$ at the edges, depending on the termination (see App.~\ref{sec:edge_terminations}).

On the strip geometry, the eigenstates of $\hat H_2$ are labeled by their momentum $k_x$ in the $x$-direction, and a band index $n \in \{1, \dots, 2N_y\}$.
The magnon operators in the LSWT eigenbasis are obtained through a Fourier transform and a subsequent unitary rotation:
\begin{align}
\label{eq:LSWT_diagonalization}
    a_{k_x,n}^\dagger = \frac{1}{\sqrt{N_x}} \sum_{y} \mathcal U_{ny}(k_x) \sum_x e^{-ik_x\sqrt 3 x} a_{i=x,y}^\dagger,
\end{align}
where $k_x$ lies in the first Brillouin zone $\mathrm{BZ} \equiv [0, 2\pi/\sqrt 3)$, and $N_x$ is the number of sites in the periodic $x$-direction.
In terms of these new operators, $\hat H_2$ is diagonal
\begin{align}
    \hat H_2 &= \sum_{k_x,n} \epsilon_{k_x,n} a_{k_x,n}^\dagger a_{k_x,n},
\end{align}
and $\epsilon_{k_x,n}$ denotes the eigenenergies.

The higher terms in the large-$S$ expansion in Eq.~\eqref{eq:large_S_expansion}, i.e., $\hat H_3$, $\hat H_4$, \textit{etc.}, encode many-body magnon-magnon interactions beyond LSWT.
Particularly interesting are the three-magnon couplings in $\hat H_3$, which are not number-conserving and facilitate spontaneous magnon decay.
These couplings are introduced if the magnetization is tilted with respect to the DMI vector, i.e., $0<\theta<\pi$, which leads to terms such as $S^{\pm} S^\parallel$ in the spin Hamiltonian that violate magnon number conservation in the Holstein--Primakoff mapping.
For our model, we find concretely
\begin{align}
\label{eq:cubic_hamiltonian_real_space}
    \hat H_3 = \frac{D_\perp \sqrt S}{\sqrt 2} \sum_{\nnn{ij}} i a_i^\dagger a_j^\dagger (a_j - a_i) + \hc
\end{align}
Note also that $\hat H_2$ depends only on $D_\parallel$ and $\hat H_3$ only on $D_\perp$.
Thus, in the following analysis, we can tune the size of the topological band gap and the cubic interaction strength independently by adjusting the magnetization angle $\theta$ and the strength $D$ of the DMI. 
It is useful to express $\hat H_3$ in terms of the eigenmodes of $\hat H_2$:
\begin{align}
\label{eq:cubic_hamiltonian_momentum_space}
    \hat H_3 &= \frac{D_\perp \sqrt S}{\sqrt{N_x}} \sumBZ{k_x,q_x} g_{k_x q_x}^{nml} a_{k_x-q_x,n}^\dagger a_{q_x,m}^\dagger a_{k_x,l} + \hc,
\end{align}
where $k_x,q_x$ and $n,m,l$ are momenta and band indices, respectively. The vertex factors $g_{k_x q_x}^{nml}$ are listed in Eq.~\eqref{eq:cubic_vertex_factors} in App.~\ref{app:magnon_magnon_vertex}.
The interaction $\hat H_3$ encodes two different number-non-conserving many-body processes: decay of one magnon into two ($\sim a^\dagger a^\dagger a$), and the Hermitian conjugate process ($h.c.$).

The higher-order terms in the large-$S$-expansion are less important for our following analysis, where we are most interested in understanding processes that can lead to magnon decay. The quartic $\hat H_4$ (and general even-magnon contributions) only contain number-conserving interactions $\sim a^\dagger a^\dagger a a$ and cannot lead to spontaneous decays. The next-leading odd contribution $\hat H_5$ does contain number-non-conserving interactions, but is suppressed by a factor $1/S$ compared to $\hat H_3$ in the large-$S$ limit.
Hence, we expect $\hat H_3$ to yield the dominant number-non-conserving correction to LSWT and correspondingly restrict our analysis to this level of the large-$S$ expansion.

\subsection{Many-Body Perturbation Theory}
\label{subsec:many_body_pert_th}

Our main goal is to study how the magnon-magnon interactions renormalize the LSWT magnon spectrum, especially the edge modes.
The magnon-magnon interactions $\hat H_{\ge 3}$ are of lower order in $S$ than $\hat H_2$ and so we treat them as a perturbation to the LSWT using the standard diagrammatic techniques.
The primary quantity of interest is the retarded single-magnon self-energy $\SE(\omega, k_x)$.
Since we consider the model on a strip geometry, $\SE(\omega, k_x)$ is a $2N_y \times 2N_y$-matrix in band space; we label its elements as $\SE_{nm}(\omega, k_x)$ with band indices $n$ and $m$.
Our main focus is on the role of \emph{spontaneous} magnon decay due to interactions. Therefore, we evaluate the self-energy at zero temperature.

In the perturbative regime, the dominant self-energy diagrams occur at one-loop order $\mathcal O(1/S^0)$.
They are simple bubbles, either made up of two cubic vertices or of one quartic vertex.
Since the quartic vertex $\hat H_4$ conserves magnon number, it is frozen out at zero temperature.
Therefore, we can omit the quartic vertex bubble.
We are left with only one dominant diagram containing two cubic vertices, which is shown in Fig.~\ref{fig:lattice_details}(d).
The corresponding self-energy at one-loop order is given by
\begin{align}
\label{eq:self_energy_one_loop}
    \SE_{nm}(\omega, k_x) = \frac{2 D_\perp^2 S}{N_x} \sum_{q_x, ab} \frac{g_{k_x q_x}^{ban} \br{g_{k_x q_x}^{bam}}^*}{\omega - \epsilon_{q_x,a} - \epsilon_{k_x-q_x,b} + i0^+},
\end{align}
where $a,b$ are intermediate band indices and the factor of $2$ is the symmetry factor of the relevant bubble diagram.
Note that the above expression is formally of order $\mathcal O(S^0)$ because the prefactor of $S$ is canceled by the linear $S$-dependence of the energies in the denominator.
From the self-energy, we compute the renormalized retarded single-magnon Green function
\begin{align}
    \GF(\omega, k_x) = \left( (\omega + i0^+) \hat{\mathbbm 1} - \hat \epsilon_{k_x} - \SE(\omega, k_x)\right)^{-1},
\end{align}
where $\hat{\mathbbm 1}$ is a $2N_y \times 2N_y$-unit matrix, $\hat \epsilon_{k_x} = \diag(\epsilon_{k_x,n})$ is a diagonal matrix of the $2N_y$ LSWT band energies at momentum $k_x$, and $\GF$ is a $2N_y \times 2N_y$-matrix in band space like the self-energy.
The poles of $\GF$, i.e., those frequencies $\tilde\epsilon_{k_x,n}$ where $\GF^{-1}$ is not invertible, are the renormalized single-magnon resonances.
Note that in the absence of magnon-magnon interactions, the poles of the propagator are simply the LSWT energies.
We can assign a band index $n \in \{1, \dots, 2N_y\}$ to each pole $\tilde\epsilon_{k_x,n}$ because every pole of the interacting propagator is uniquely (up to degeneracy) connected with a pole of the bare (LSWT) propagator by adiabatically turning the magnon-magnon interaction strength $D_\perp$ to zero.

Both the self-energy and the single-magnon propagator capture how the interactions couple the single-magnon modes to the two-magnon continuum.
Specifically, they encode the following renormalization effects:
\begin{enumerate}
    \item \textit{Decay:} 
    The anti-Hermitian part of the self-energy, $(\SE - \SE^\dagger)/2$, gives the single-magnon modes a finite lifetime as a single magnon can irreversibly decay into the two-magnon continuum.
    In particular, an edge magnon can decay into two bulk magnons, potentially suppressing well-defined long-lived edge excitations. In technical terms, decay effects shift the poles of the Green function to complex values; their imaginary parts are proportional to the respective \emph{magnon decay rates}:
    \begin{align}
        \Gamma_{k_x,n} = -2\im\tilde\epsilon_{k_x,n}
    \end{align}
    where the factor of $2$ is convention \footnote{With a factor of $2$, $\Gamma$ represents the decay rate of the physical probability of a single magnon state according to Fermi's golden rule, rather than the decay rate of the probability amplitude.}.
    In Sec.~\ref{sec:spontaneous_decay_rates}, we explain in detail how we numerically quantify magnon decay rates.
    
    \item \textit{Dispersion renormalization:} The Hermitian part of the self-energy, $(\SE + \SE^\dagger)/2$, shifts the real part of the Green function poles away from the LSWT dispersion.
    This effect can be understood as a level repulsion of the single-magnon modes from the two-magnon continuum \cite{verresen_avoided_2019}.
    We call $\re\tilde\epsilon_{k_x,n}$ the \emph{renormalized magnon dispersion}.
    
    \item \textit{Hybridization:} A single magnon in band $n$ can decay into an intermediate pair of virtual magnons, which may then recombine to yield a single magnon in a different band $m \neq n$.
    This leads to non-zero off-diagonal self-energy terms $\SE_{n\neq m}(\omega, k_x)$, and the renormalized single-magnon wavefunctions are hybrids of different LSWT band eigenstates \cite{mook_interaction-stabilized_2021} as analyzed in detail in Sec.~\ref{sec:spontaneous_deloc_resonances}.
\end{enumerate}

\section{Renormalized Single-Magnon Spectra}
\label{sec:renormalized_spectra}

\figTwoColumn{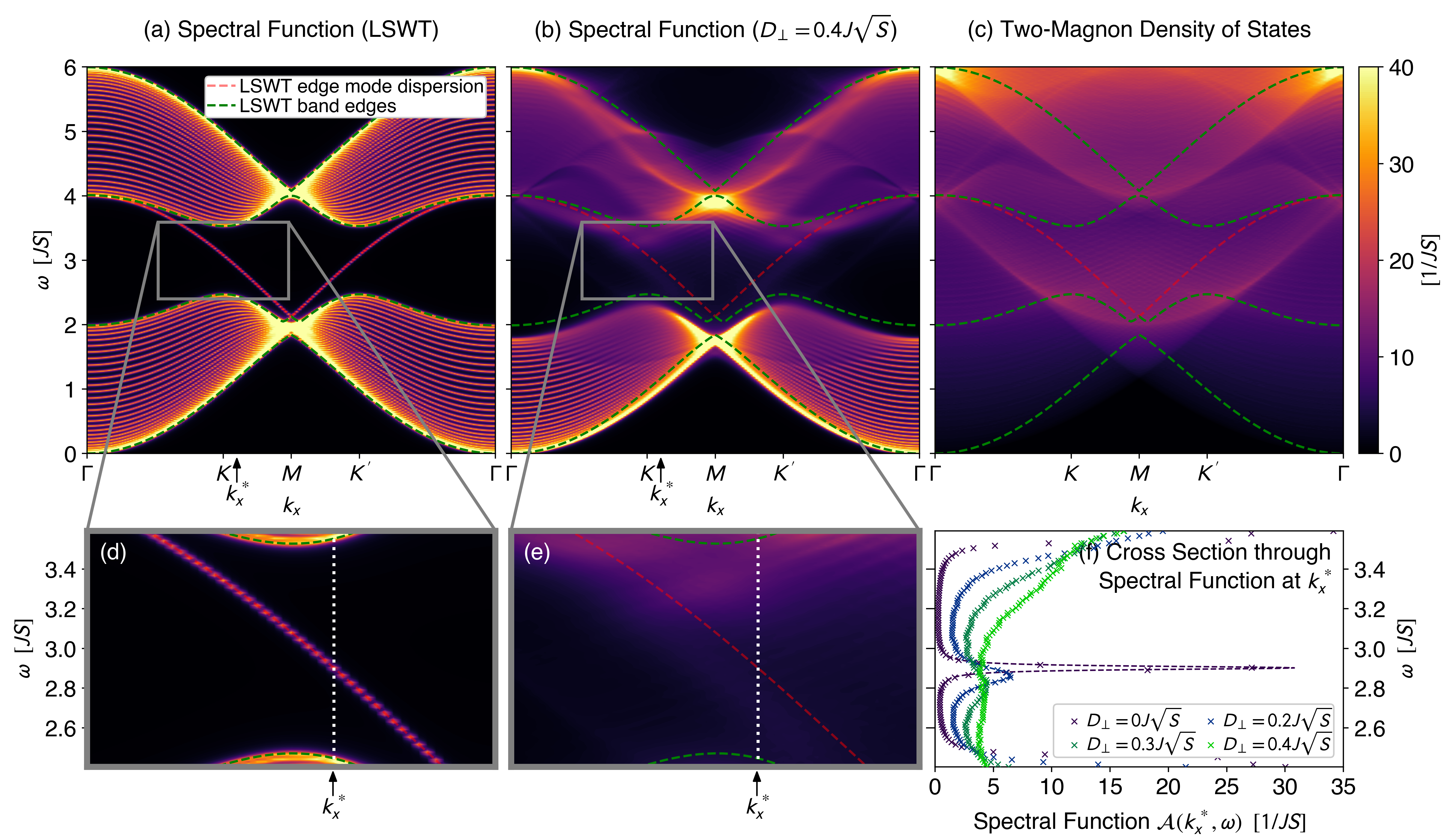}{\captionSpectraZigzag}
    
To analyze the single-magnon spectrum in the presence of many-body interactions, we numerically compute the spectral function
\begin{align}
\label{eq:spectral_function}
    \SF(\omega, k_x) &\equiv -\frac{1}{\pi} \im\tr \GF(\omega, k_x).
\end{align}
It is a useful diagnostic tool because it is closely related to observables like the dynamical spin structure factor, which can be probed experimentally, e.g., by inelastic neutron scattering.
If a single-magnon mode has an infinite lifetime, it appears as a sharp delta peak in the spectrum at the corresponding renormalized magnon dispersion $\re\tilde\epsilon_{k_x,n}$, although its spectral weight can be reduced by a quasiparticle weight $Z_{k_x,n} \le 1$.
Decay effects, such as those 
induced by many-body interactions coupling the single- and two-magnon sectors, broaden this peak into a Lorentzian whose width is exactly the decay rate $\Gamma$. Therefore, the formerly sharp single-magnon response becomes an unstable resonance.
In addition to this broadened but ``coherent" single-magnon resonance, there is a continuum $\SF_\mathrm{incoh}$ of incoherent background spectral weight that cannot be associated with a well-defined single-magnon excitation.
In the vicinity of a pole $\tilde\epsilon_{k_x,n}$, we can therefore write the spectral function as
\begin{align}
    \SF(\omega, k_x) = \frac{Z_{k_x,n} \Gamma_{k_x,n}/2}{(\omega - \re\tilde\epsilon_{k_x,n})^2 + (\Gamma_{k_x,n}/2)^2} + \SF_\mathrm{incoh}(\omega, k_x).
\end{align}

In the following numerical results, we consider a strip with $N_y = 30$ unit cells in the non-periodic $y$-direction \footnote{$N_y=30$ unit cells is large enough to prevent LSWT-level hybridization of the edge modes at opposite sites of the strip in the numerics; this is because the LSWT hybridization gap decays exponentially for strip sizes larger than the typical localization lengths of the edge modes, which are only a few unit cells.} and set the DMI compoment parallel to the magnetization to $D_\parallel = 0.1JS$ unless stated otherwise.

In Fig.~\ref{fig:spectra_zigzag}, we present two full single-magnon spectra for the zig-zag edge termination with and without magnon-magnon interactions.
Without interactions [$D_\perp = 0.0 J \sqrt S$ in Fig.~\ref{fig:spectra_zigzag}(a)], the spectrum corresponds exactly to the LSWT band structure, with delta peaks in frequency space centered at the band energies.
In particular, the chiral edge modes inside the band gap are sharp and thus stable, as predicted by the LSWT bulk-boundary correspondence [see Fig.~\ref{fig:spectra_zigzag}(d)].

In the presence of interactions [$D_\perp = 0.4 J\sqrt S$ in Fig.~\ref{fig:spectra_zigzag}(b)], however, the spectrum is significantly renormalized \footnote{$D_\perp = 0.4 J\sqrt S$ is the maximal interaction strength that ensures a ferromagnetic ground state for reasonable $S$ and the validity of the perturbative treatment, see App.~\ref{app:validity_pert_th}}.
The upper half of the band structure is washed out by life-time broadening, and the bands tend to be pushed downwards as a result of the level repulsion between single-magnon bands and the two-magnon continuum.

Most strikingly, the spectral peaks of the edge modes vanish completely [see Fig.~\ref{fig:spectra_zigzag}(e)].
Instead, a sizable incoherent spectral background continuum emerges inside the bulk gap as spectral weight is drawn into the gap.
As can be seen in Fig.~\ref{fig:spectra_zigzag}(f), this process happens gradually for increasing interaction strength.
At $D_\perp = 0.3J\sqrt S$, we still observe a weak Lorentzian resonance inside the band gap, albeit severely broadened and offset by incoherent background.
At $D_\perp = 0.4J\sqrt S$, this resonance is completely overshadowed by the incoherent background and thus becomes invisible in the spectrum.

We can understand why the renormalization effects are so dramatic by inspecting the two-magnon density of states [see Fig.~\ref{fig:spectra_zigzag}(c)]:
it is fairly weak at low frequencies, but jumps abruptly to relatively high values in the region of the LSWT band gap.
This jump appears due to the opening of a new decay channel: an edge magnon is suddenly kinematically allowed to decay into two final bulk magnons, one of which lives at the almost degenerate $M$-point.
This strongly increases the available decay phase space, enhancing the lifetime broadening of the edge modes and the accumulation of incoherent background inside the band gap.

Our main result is that magnon-magnon interactions can severely alter the topological features predicted by LSWT; in particular, the bulk-boundary correspondence from LSWT may not accurately predict the edge physics of the full spin model.
On a conceptual level, the incoherent background inside the LSWT band gap gives rise to a continuum single-magnon-like states, which effectively closes the gap invalidating the computation of Chern numbers as these require gapped bands.

\figTwoColumn{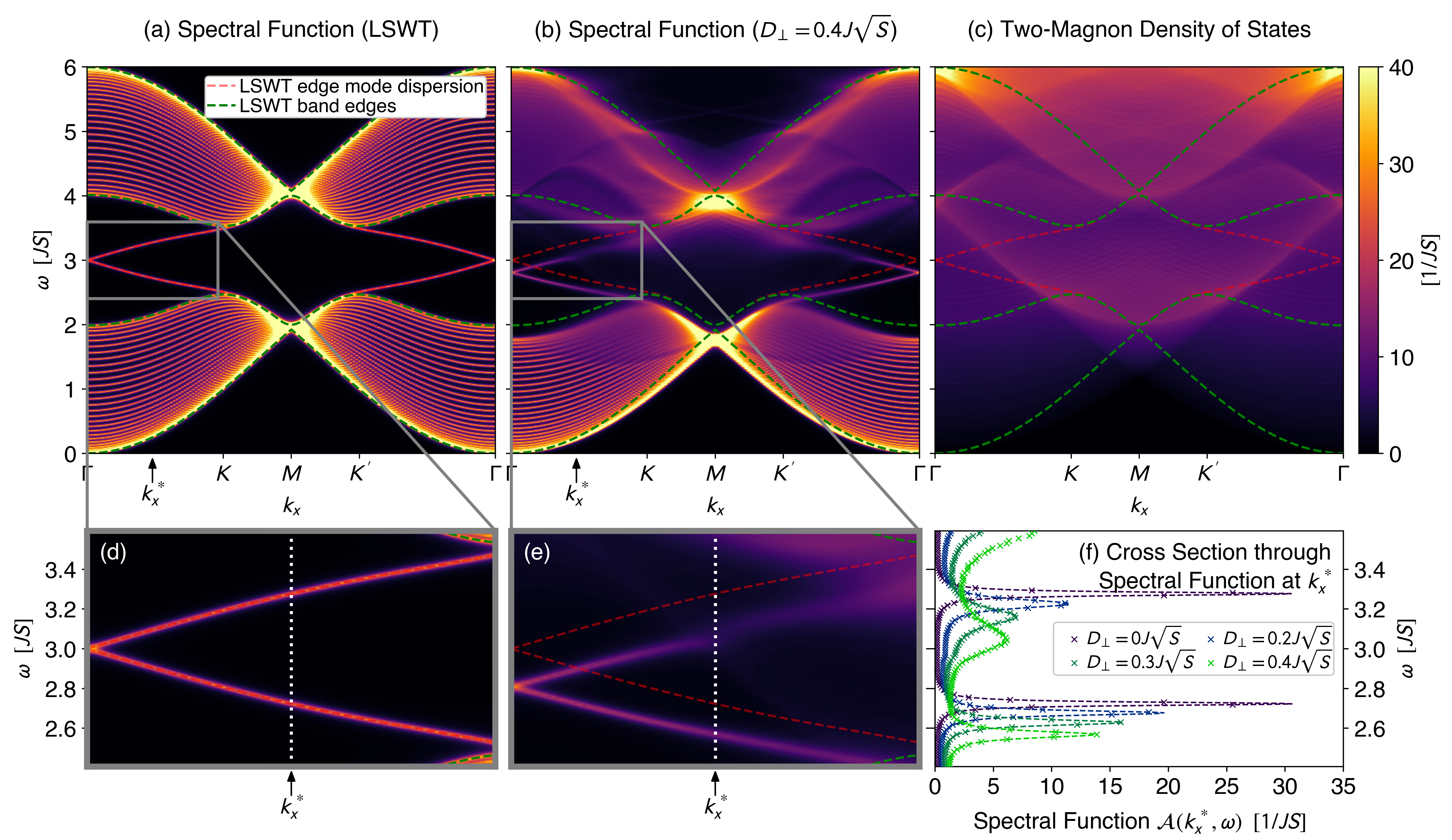}{\captionSpectraDangling}

The situation is less dramatic for the strip with dangling edge termination (see Fig.~\ref{fig:spectra_dangling}).
Even for the strongest interaction $D_\perp = 0.4 J\sqrt S$, the edge modes appear as clearly distinguishable resonances in the spectrum, albeit in part heavily broadened [see Figs.~\ref{fig:spectra_dangling}(e) and (f)].
The degree of broadening is correlated with the two-magnon density of states: the lower-energy edge mode lies in a region of low two-magnon density of states and thus remains relatively sharp; the higher-energy edge mode suddenly acquires a more pronounced broadening at $k_x^*$, where it crosses the aforementioned step in the two-magnon density of states, and decays to the $M$-point become kinematically possible.
Nonetheless, the lifetime broadening of the edge modes indicates that their renormalized wavefunctions generally contain admixtures from bulk-like two-magnon states.
Hence, the single-magnon picture from LSWT again fails to describe the magnon physics accurately in the presence of magnon-magnon interactions.

In general, we observe that the impact of magnon-magnon interactions on the topological edge modes is non-universal:
while the existence of edge modes is topologically guaranteed in LSWT, they may be absent beyond LSWT, depending on microscopic details like the edge termination.
This non-universality is due to the high-energy nature of the magnon edge modes, which makes them prone to decay.
This behavior is very unlike electronic edge modes near the Fermi energy, which are universally protected against decay by a lack of final decay phase space at low-energies.

\section{Spontaneous Decay Rates of Edge Modes}
\label{sec:spontaneous_decay_rates}

\fig{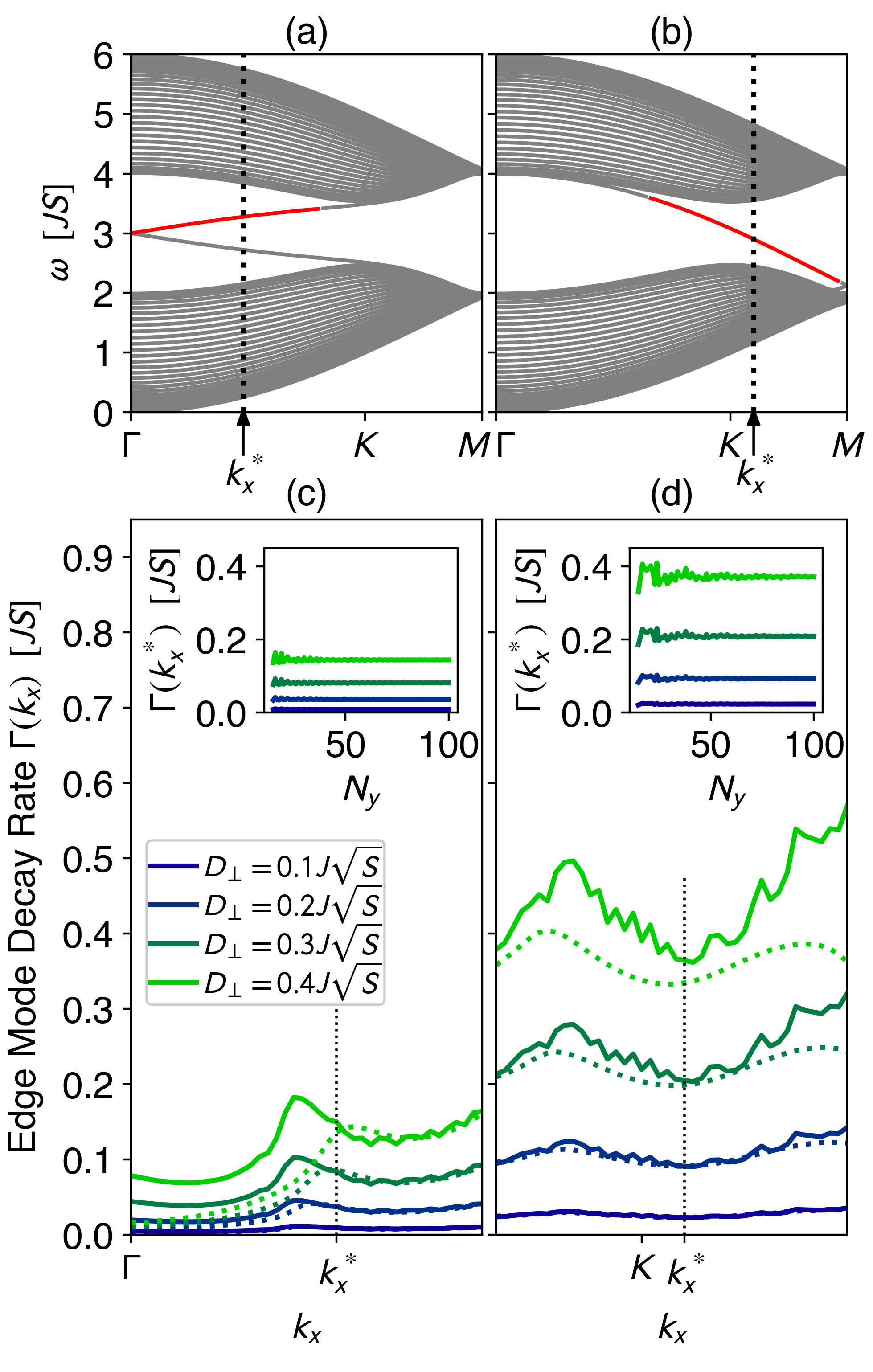}{\captionDecayRates}

To quantify the life-time broadening of the edge modes, we can extract their decay rates $\Gamma_{k_x,n} = -2\im\tilde\epsilon_{k_x,n}$, where $n$ is the band index of the edge mode of interest.
To obtain $\tilde\epsilon_{k_x,n}$, one must self-consistently solve the pole condition
\begin{equation}
\label{eq:pole_equation}
    \det\left((\tilde\epsilon_{k_x,n} + i0^+) \hat{\mathbbm 1} - \hat\epsilon_{k_x} - \SE(\tilde\epsilon_{k_x,n}^*, k_x)\right) = 0,
\end{equation}
where $(\dots)^*$ is complex conjugation \cite{chernyshev_spin_2009}.
This is numerically cumbersome, especially for the system sizes which we investigate.
To circumvent these difficulties, an established way to estimate the decay rates is the on-shell approximation \cite{zhitomirsky_colloquium_2013} (for a derivation, see App.~\ref{app:decay_rate_os_approx}),
\begin{align}
    \Gamma_{k_x,n}^\mathrm{os} = -2 \im \Sigma_{nn}(\epsilon_{k_x,n}, k_x),
\end{align}
where $\epsilon_{k_x,n}$ is the bare (``on-shell") energy of mode $n$ obtained from LSWT.
This simple and quick approximation is exact up to order $\mathcal O(1/S^0)$, but it can neither capture hybridization of LSWT bands (since it disregards the off-diagonal self-energy elements) nor dispersion shifting due to level repulsion (since it does not know about the real part of the self-energy).

Here, we use both a self-consistent solution to the pole condition and the on-shell approximation.
We later compare the two methods to assess the validity of the latter (also cf.\ Sec.~\ref{sec:magnetic_fields}).
Figure~\ref{fig:decay_rates_edge_mode} shows the self-consistent (dotted lines) and on-shell (dashed lines) decay rates of a selected edge mode versus momentum $k_x$ for dangling [LSWT spectrum in Fig.~\ref{fig:decay_rates_edge_mode}(a), decay rates in Fig.~\ref{fig:decay_rates_edge_mode}(c)] and zig-zag edge terminations [LSWT spectrum in Fig.~\ref{fig:decay_rates_edge_mode}(b), decay rates in Fig.~\ref{fig:decay_rates_edge_mode}(d)].
In the case of dangling edges, we find that the decay rates of the higher-energy edge mode show a peak close to the reference momentum $k_x^*$ where the edge mode dispersion enters a region of higher two-magnon density of states (as detailed in Sec.~\ref{sec:renormalized_spectra}).
We note that this peak appears at a higher momentum in the self-consistent decay rates as compared to the on-shell values.
This is because the self-consistent treatment naturally takes into account level repulsion of the edge mode from the two-magnon continuum (which allows it to stay in regions of low two-magnon density of states up to higher momenta), while the on-shell approximation ignores this effect.
Yet, we observe that even for the strongest magnon-magnon interaction strength $D_\perp = 0.4 J \sqrt S$, the peak decay rate is only about $0.2 JS$, much smaller than the topological band gap of approximately $1JS$ at the $K$-point.
Therefore, the edge resonances are broadened but still clearly visible.

For the zig-zag edge termination, the decay rates are generally much higher than for the dangling edges.
For $D_\perp = 0.4 J \sqrt S$, they reach up to about $0.5 JS$, which is of the order of magnitude of the topological band gap around the $K$-point.
In the presence of such severe lifetime broadening, speaking of a well-defined mode inside the band gap is questionable.

As a side note, the on-shell decay rates in Figs.~\ref{fig:decay_rates_edge_mode}(c) and (d) appear slightly jagged, which we attribute to finite-size effects: the finite number of lattice sites in $y$-direction leads to van-Howe singularities in the two-magnon density of states originating from the quasi-1D $k_x$ dispersion, which are immediately reflected in the on-shell decay rate. 
In contrast, the self-consistent decay rates intrinsically take into account the lifetime broadening of the decaying edge magnon so that the van-Howe singularities are smoothed out.
We emphasize that in the limit of $N_y \to \infty$, the jaggedness of the on-shell decay rates disappears and the decay rate converges to a constant, as shown in the insets of Figs.~\ref{fig:decay_rates_edge_mode}(c) and (d).

\section{Spontaneous Hybridization}
\label{sec:spontaneous_deloc_resonances}

\subsection{Edge-Bulk Hybridization}
\label{subsec:edge_bulk_hybridization}

\fig{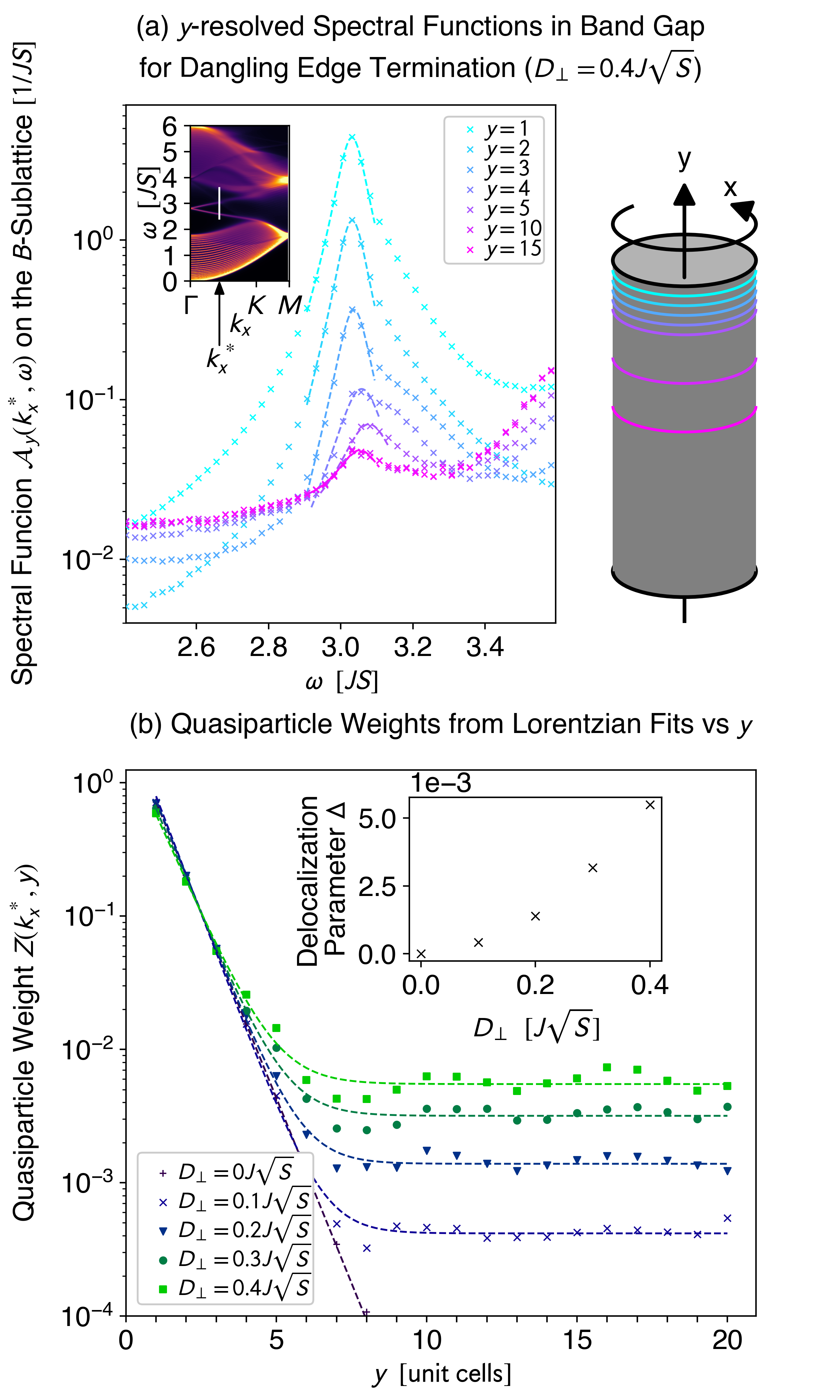}{\captionSpectraRealSpace}

Beyond the study of decay rates, off-diagonal self-energy terms produce renormalized hybrids of different LSWT eigenstates (see Sec.~\ref{subsec:many_body_pert_th}).
Most remarkably, we find that the edge modes can hybridize with delocalized bulk states.
Because this already happens at zero temperature, \emph{spontaneous} edge-bulk hybrid resonances form.

In order to investigate these delocalized hybrid resonances, we first consider a modification of the spectral function which resolves the real-space $y$-coordinate \cite{mook_interaction-stabilized_2021}:
\begin{align}
\label{eq:y_resolved_spectral_function}
    \SF_y(\omega, k_x) \equiv -\frac{1}{\pi} \im \br{\sum_{nm} \mathcal U_{yn}^\dagger(k_x) \GF_{nm}(\omega, k_x) \mathcal U_{my}(k_x)}.
\end{align}
Here, $n,m$ are band indices, and the matrix elements $\mathcal U_{yn}(k_x)$ constitute the unitary transformation that diagonalizes the $k_x$-space blocks of the LSWT Hamiltonian $\hat H_2$ as defined in Eq.~\eqref{eq:LSWT_diagonalization}.
Note that summing this $y$-resolved spectral function over all $y$ recovers the standard spectral function in Eq.~\eqref{eq:spectral_function}, as $\mathcal U(k_x)$ is unitary:
\begin{align}
    \sum_y \SF_y(\omega, k_x) = \SF(\omega, k_x).
\end{align}

In the following, we will only investigate the case of dangling edge terminations since the edge modes for the zig-zag edges are severely overdamped.
We will also focus on the reference momentum $k_x^*$ (about halfway between the $\Gamma$ and $K$-point) as indicated in Fig.~\ref{fig:spectra_dangling}: at this momentum, the two-magnon density of states has a step due to the opening of a new decay channel to the $M$-point.
Since the edge-bulk hybridization is mediated by virtual two-magnon states, we expect that it is particularly pronounced around this momentum.

In Fig.~\ref{fig:spectra_real_space}(a), we plot the $y$-resolved spectral function versus the frequency window of the band gap for different values of $y$ with strong magnon-magnon interactions $D_\perp = 0.4J\sqrt S$.
For small $y$ close to the lower edge of the strip, we see clear Lorentzian resonances stemming from the higher-energy LSWT edge mode; the amplitudes of these Lorentzians decrease for increasing $y$.
Without magnon-magnon interactions, we would expect them to become exponentially small when reaching values of $y$ deep inside the bulk.
With magnon-magnon interactions, we instead observe that the resonances do not tend to zero but there remains a residual peak even at $y = N_y/2 = 15$ in the middle of the bulk.
This is a signature that the interactions admix the edge modes with delocalized bulk states.

Next, we fit Lorentzian shapes to the resonances in Fig.~\ref{fig:spectra_real_space}(a) to extract their $y$-resolved quasiparticle weight $Z_y$.
Figure~\ref{fig:spectra_real_space}(b) shows these quasiparticle weights versus $y$ for different interaction strengths $D_\perp$.
They decay exponentially near the edge, but then level off at larger $y$ and approach a finite value which depends on the interaction strength.
For each $D_\perp$, we also fit a heuristic function $A e^{-2y/L} + \Delta$ to the data, where $A$ is a constant, $L$ corresponds to the localization length of the edge resonance, and $\Delta$ is a \emph{delocalization parameter} which measures how much of the spectral weight near the edge mode energy is contributed by bulk states. 
In the non-interacting case $D_\perp = 0$, we observe a pure exponential decay of the quasiparticle weight with $\Delta = 0$, in accordance with the LSWT prediction of a localized edge mode.
Including magnon-magnon interactions, however, $\Delta$ grows quadratically with $D_\perp$ [see inset of Fig.~\ref{fig:spectra_real_space}(b)].

To understand the behavior of $\Delta$, we perform a perturbative analysis (see App.~\ref{app:deloc_param_perturbative}).
We find that the interactions mix single-magnon bulk states and delocalized two-magnon states into the renormalized edge mode wavefunction.
As a consequence, the $y$-resolved spectral function undergoes corrections at order $\mathcal O(D_\perp^2)$.
The corrections' delocalized nature becomes apparent in the bulk tails of the quasiparticle weights, as demonstrated in Fig.~\ref{fig:spectra_real_space}(b), explaining the observed scaling of $\Delta$.
We note, however, that the $y$-resolved spectral function is unable to distinguish clearly between delocalized contributions from single- and two-magnon states.

\fig{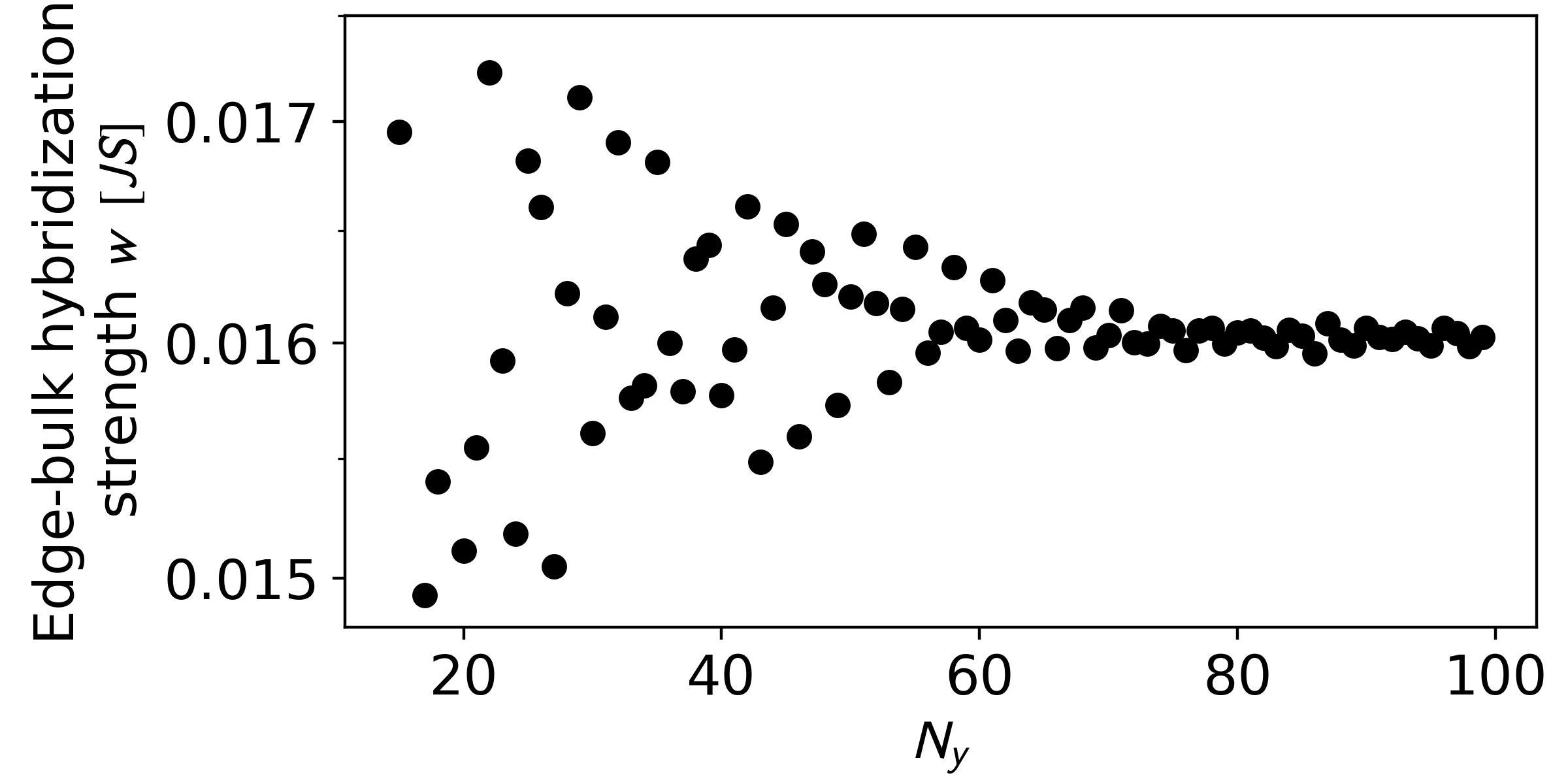}{\captionEdgeBulkOverlapVsSlabSize}

Therefore, to highlight specifically the hybridization strength between single-magnon edge and bulk states, we perturbatively compute the squared overlap of the renormalized edge mode at the lower edge $\ket\psi$ with all single-magnon bulk states (see App.~\ref{app:deloc_param_perturbative}):
\begin{align}
    w \equiv \sum_{m\in\mathrm{bulk}} \left| \braket{0 | a_{k_x^*,m} | \psi} \right|^2 = \sum_{n\in{\mathrm{bulk}}} \left| \frac{\SE_{ln}(\epsilon_{k_x^*,l}, k_x^*)}{\epsilon_{k_x^*,l} - \epsilon_{k_x^*,n}} \right|^2,
\end{align}
where $l$ is the band index of the edge mode localized at the lower edge, and $\sum_{m\in\mathrm{bulk}}$ is the sum over bulk band indices.
Fig.~\ref{fig:edge_bulk_overlap_vs_slab_size} shows $w$ versus strip width $N_y$ at interaction strength $D_\perp = 0.4J\sqrt S$.
The plot suggests that $w$ approaches a constant value of approximately 0.016 in the thermodynamic limit $N_y \rightarrow \infty$.
Hence, the renormalized edge mode is a hybrid state that contains 1.6\% single-magnon bulk states, resulting in a delocalized wavefunction that is non-zero even far away from the edges.

In conclusion, we find that magnon-magnon interactions are capable of hybridizing the localized edge modes from LSWT with delocalized bulk modes even at $T=0$.
This is further evidence that the LSWT bulk-boundary correspondence gives inaccurate predictions about the edge physics in the full spin model where magnon-magnon interactions are important.

\subsection{Edge-Edge Hybridization}
\label{subsec:edge_edge_hybridization}

\fig{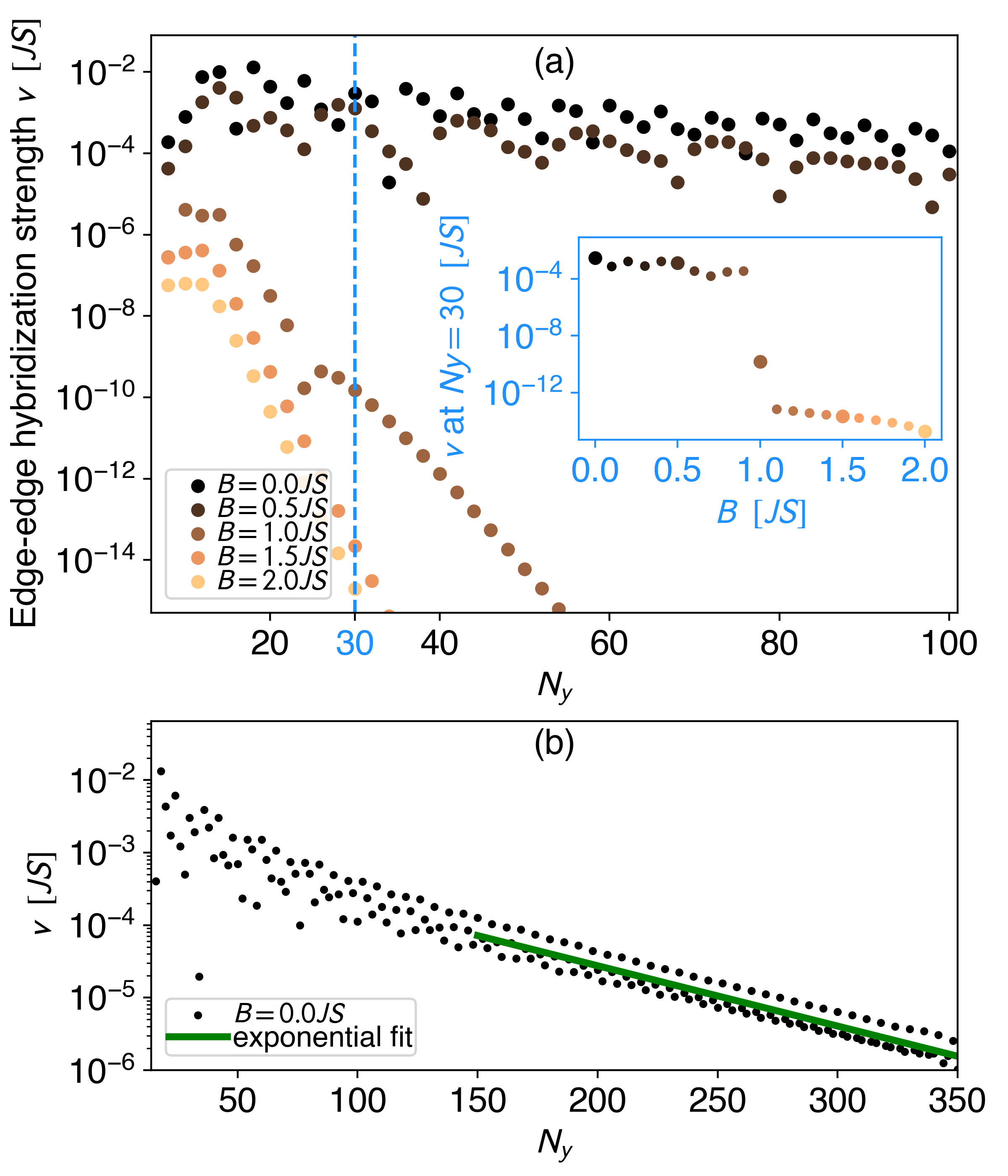}{(a) Edge-edge hybridization strength $v$ versus number of unit cells in $y$-direction for different magnetic fields at $D_\perp = 0.4J\sqrt S$ (only even $N_y$ are considered).
The inset shows $v$ versus a higher-resolution range of magnetic field values at fixed $N_y=30$ (indicated by the dashed blue line in the main plot).
It showcases the sudden drop of the edge-edge hybridization at $B=1.0JS$ connected to the separation of the LSWT edge modes and the two-magnon continuum.
The data points in the inset are bold when the corresponding magnetic field is included in the main plot.
Panel (b) shows $v$ versus $N_y$ for zero magnetic field on a semi-log scale. An exponential is fitted in the interval between $N_y=150$ and $350$; its decay length is about $52$ unit cells, much larger than the LSWT edge mode localization length of $0.97$ unit cells.
}

Besides the hybridization of edge modes with bulk states, magnon-magnon interactions may also mediate hybridization \emph{between} edge modes of opposite chirality and support on opposite sides of the strip.
This effect is particularly relevant at the crossing points of two edge modes: at these points, they become energetically degenerate so that an arbitrarily small coupling can give rise to an avoided crossing and formation of edge-edge hybrid states.
In our honeycomb TMI on the strip, we expect the edge-edge hybridization to be more pronounced for the zig-zag edge termination than for the dangling edge termination because the edge mode crossing for the former lies at the $M$-point in a region of high two-magnon density of states, while the crossing for the latter is at the $\Gamma$-point where two-magnon states are relatively scarce. 
Therefore, we only consider the zig-zag edges in the following.
We can describe the interaction-mediated edge-edge coupling by means of an effective non-Hermitian on-shell Hamiltonian in the subspace of the two crossing edge modes.
It is composed of the LSWT Hamiltonian and the self-energy matrix at the momentum of the crossing point $k_x=M$, evaluated at the corresponding LSWT (on-shell) crossing energy $\epsilon_{\times}$ and projected into the edge mode subspace.
Labeling the band indices of the edge modes localized at the upper and lower edge $u$ and $l$, the effective Hamiltonian reads
\begin{align}
    \hat H_\mathrm{eff} = \begin{pmatrix}
        \epsilon_\times + \SE_{uu}(\epsilon_{\times}, M) & \SE_{ul}(\epsilon_{\times}, M) \\
        \SE_{lu}(\epsilon_{\times}, M) & \epsilon_\times + \SE_{ll}(\epsilon_{\times}, M)
    \end{pmatrix},
\end{align}
where $\epsilon_{\times} = \epsilon_{M,u} = \epsilon_{M,l}$.
In general, the anti-Hermitian part of $\hat H_\mathrm{eff}$ may give rise to exceptional points \cite{mcclarty_non-hermitian_2019}.
However, neglecting these complications and only focusing on the Hermitian part, we can read off the interaction-induced edge-edge hybridization strength to be $v \equiv \left|\SE_{ul}(\epsilon_\times, M) + \SE_{lu}(\epsilon_\times, M)\right| / 2$.

Note that, in principle, the edge modes also hybridize at LSWT level for sufficiently thin strips due to the closeness of the edges.
The thus induced hybridization gap decays exponentially with the strip width $\sim \exp(-N_y/L)$, where $L$ is the typical LSWT localization length of the edge modes.
At the $M$-point, $L$ is only about $0.97$ unit cells, so the hybridization at LSWT level is already insignificant for very thin strips.
In contrast, the interaction-induced edge mode coupling $v$ is fundamentally different as it is mediated by delocalized two-magnon states.
To demonstrate how $v$ scales with strip size, we plot the numerical values versus a range of strip widths between $N_y = 8$ and $100$ at the strongest probed interaction strength $D_\perp = 0.4J\sqrt S$ [see Fig.~\ref{fig:hybridization_vs_slab_size}(a)].
In this section, we only focus on the case of zero magnetic field ($B=0.0JS$, black dots) and defer the discussion of the non-zero field data to Sec.~\ref{sec:magnetic_fields}.
Note that, in Fig.~\ref{fig:hybridization_vs_slab_size}, we only include even $N_y$; for odd $N_y$, a subtle cancellation effect in the self-energy matrix elements renders the hybridization strength several orders of magnitude lower than the values for even $N_y$.

At zero magnetic field, we observe a very slow decay of $v$ with the strip width.
Like in the LSWT limit, it is still exponential for larger strip sizes as depicted in Fig.~\ref{fig:hybridization_vs_slab_size}(b).
However, an exponential fit reveals a decay length of about $52$ unit cells -- almost two orders of magnitude larger than the LSWT edge mode localization length of $0.97$ unit cells.
Although the numerical data suggest that $v$ tends to zero in the thermodynamic limit $N_y\rightarrow\infty$, we conclude that the edge-edge hybridization may still lead to a breakdown of topological edge mode protection in nano-scale systems.

As a side note, our numerical data indicate that also the anti-Hermitian part of $\hat H_\mathrm{eff}$ decays to zero exponentially for large $N_y$.
The corresponding decay length is comparable to that of the Hermitian part $v$.

\section{Restoring Magnon Topology with Magnetic Fields}
\label{sec:magnetic_fields}

\figTwoColumn{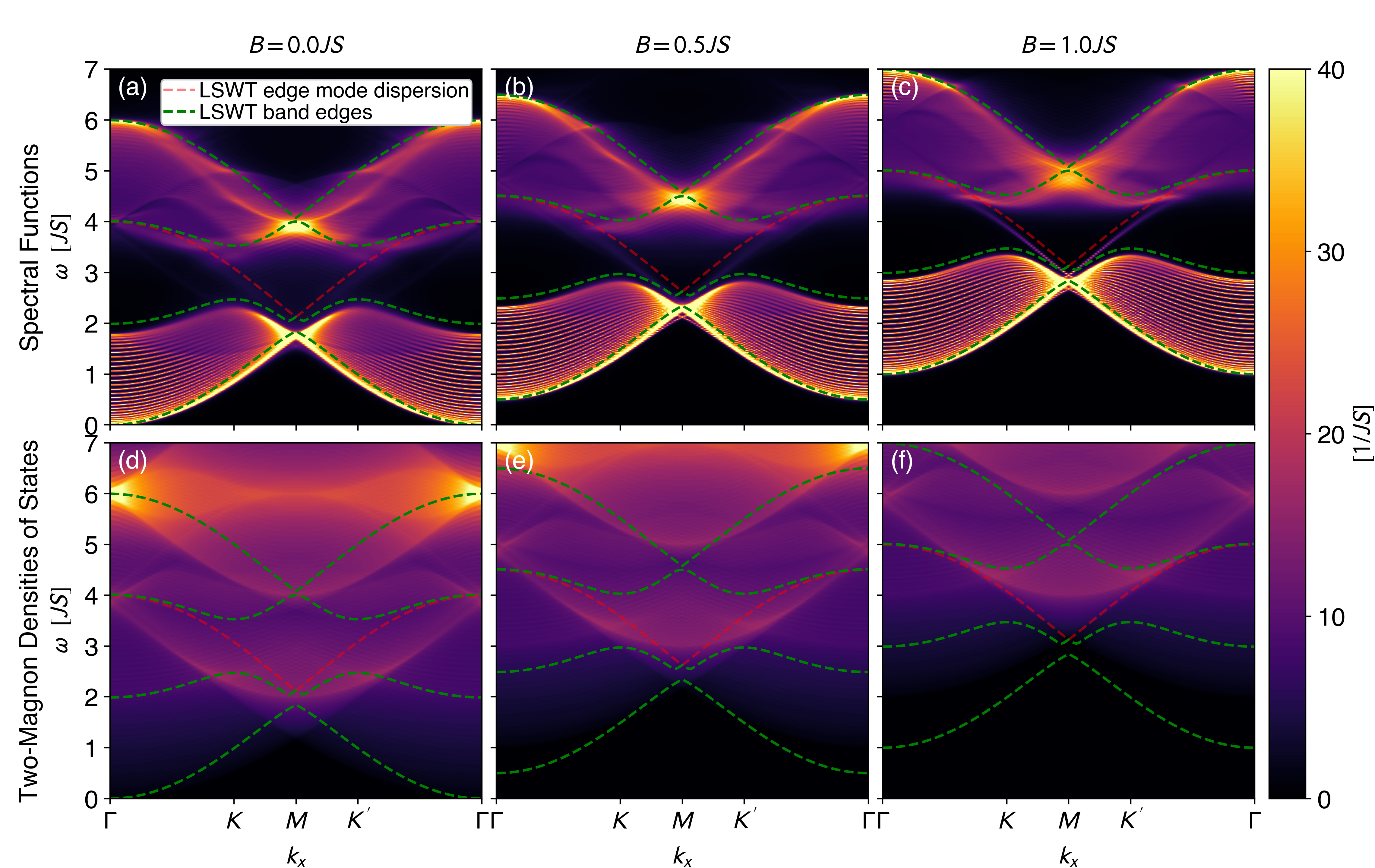}{\captionSpectraZigzagBField}

\figTwoColumn{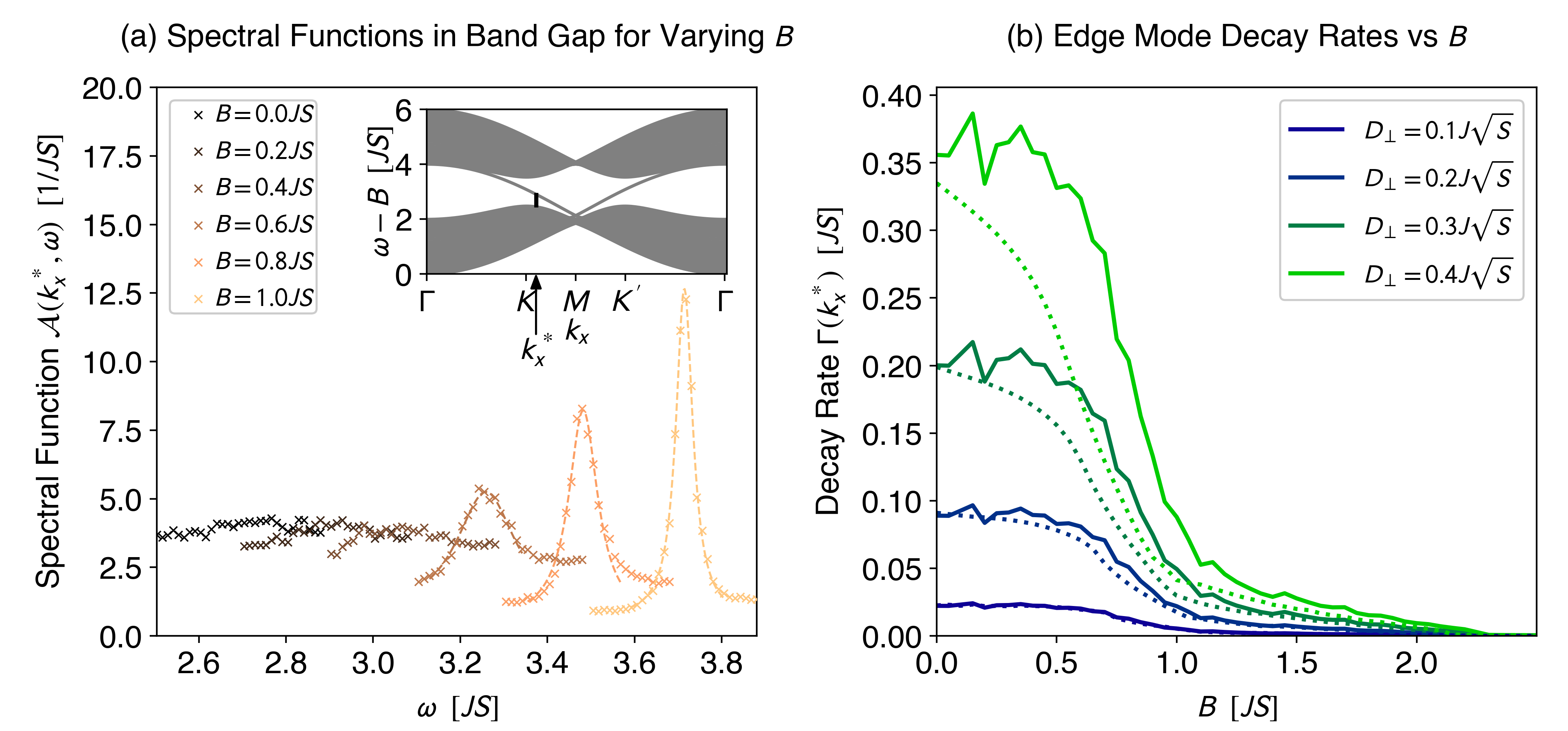}{\captionDecayRatesZigzagBField}

So far, we have demonstrated that magnon-magnon interactions may strongly affect or even completely erase topological edge modes in the single-magnon spectrum in non-universal ways.
We have focused on the situation of energetically overlapping single- and two-magnon spectra, which is the crucial ingredient to find magnon decay in perturbation theory \cite{zhitomirsky_colloquium_2013}.
Importantly, this overlap can be tuned by an external magnetic field.
Indeed, Ref.~\onlinecite{mcclarty_topological_2018} has shown that topological magnon edge modes are stable in a field-polarized generalized Kitaev-Heisenberg model at strong enough fields.
It turns out that the magnetic field, which in this model is required to have topological magnons in the first place, separates single-magnon bands and multi-magnon continua, and thus prohibits edge mode decays kinematically.
In contrast, the DMI honeycomb ferromagnet does not require a magnetic field to host topological magnons. 
Next, we demonstrate that, in a similar way, a magnetic field is able to restore sharp edge modes and the topological features from LSWT by kinematically suppressing magnon decay for the DMI honeycomb ferromagnet.

A magnetic field $\vec B$ pointing in the direction of the ground state magnetization acts like a chemical potential for the magnons.
We model it by adding a Zeeman term 
\begin{align}
    \hat H_\mathrm{Zeeman} = -\vec B \cdot \sum_i \vec S_i = - B \sum_i (S - a_i^\dagger a_i)
\end{align}
to the Hamiltonian, where $B \equiv |\vec B|$.
This term shifts up the LSWT single-magnon energies by a constant offset of $B$.
Crucially, the two-magnon continuum is also shifted, but by an offset of $2B$, since the magnetic field couples equally to both magnons.
As a consequence, the two-magnon continuum is eventually pushed past the single-magnon bands for increasing $B$, energetically freezing out the decay channel of one magnon into two, and thus stabilizing the edge modes.
The same argument can be extended to higher-order decay into $n\ge 3$ final magnons, which may be induced by the terms $\hat H_{n\ge 4}$ in the large-$S$ expansion of the Hamiltonian, or by higher powers of $\hat H_3$.
The $n$-magnon continuum required for these processes is shifted by an offset of $nB$, so they are already suppressed at even smaller $B$ than the decay into two final magnons \cite{Harris1971}.
Therefore, the high-field regime provides a limit in which the bulk-boundary correspondence from LSWT holds rigorously even in the presence of arbitrary magnon-magnon interactions, reinstalling a well-defined notion of magnon topology.

In Fig.~\ref{fig:spectra_zigzag_B_field}, we show single-magnon spectra and the corresponding two-magnon densities of states for increasing magnetic field in the presence of strong magnon-magnon interactions, $D_\perp = 0.4J\sqrt S$.
Without magnetic field, as depicted in Fig.~\ref{fig:spectra_zigzag_B_field}(a,d), the LSWT edge modes completely vanish from the spectrum because they couple to a large two-magnon density of states (as detailed in Sec.~\ref{sec:renormalized_spectra}).
As the magnetic field increases, the two-magnon density of states at the LSWT edge mode dispersion decreases while the edge resonances first become visible again [see Fig.~\ref{fig:spectra_zigzag_B_field}(b,e)] and then sharpen significantly [see Fig.~\ref{fig:spectra_zigzag_B_field}(c,f)].
Although decays are suppressed in strong magnetic fields, the entire spectrum is still subject to level repulsion from the two-magnon continuum, renormalizing the single-magnon modes downward (relative to the dashed green lines).

A quantitative analysis of the field-induced spectral reappearance of the edge resonance is shown in Fig.~\ref{fig:decay_rates_zigzag_B_field}. The spectral function is plotted for selected magnetic field strengths in Fig.~\ref{fig:decay_rates_zigzag_B_field}(a), with reference momentum $k_x^*$ and frequency window indicated in the inset.
For $B \le 0.4JS$, the spectrum is dominated by incoherent background; but at $B = 0.6JS$, an approximate Lorentzian suddenly emerges inside the band gap, signaling the restoration of a distinct edge resonance.
For even higher $B$, the resonance gets sharper due to the kinematic suppression of decays, and it is shifted to higher frequencies due to the coupling to the magnetic field.

To quantify the field-induced suppression of decays, we extract the decay rates of the discussed edge resonance at momentum $k_x^*$ as a function of $B$, using both on-shell approximation and a self-consistent solution to the pole condition.
Figure \ref{fig:decay_rates_zigzag_B_field}(b) shows the decay rates versus $B$ for different magnon-magnon interaction strengths.
For low $B$, edge magnons can decay into bulk magnons at the almost degenerate $M$-point (see Sec.~\ref{sec:renormalized_spectra}), significantly boosting the decay rates and leading to their disappearance in the incoherent background.
For $B \ge 0.5JS$, this decay channel begins to close as the two-magnon continuum is gradually shifted past the single-magnon bands, leading to a sudden strong reduction of the decay rate between $B = 0.5JS$ and $B = 1.0JS$.
Beyond $B = 1.0JS$, edge magnons can only decay to low-energy bulk magnons near the $\Gamma$-point, which are not very dense, such that the decay rates level off at very low values ($\lesssim 0.05JS$ for $D_\perp = 0.4J\sqrt S$, much smaller than the band gap).
As $B$ is increased even further, the decay rates slowly drop further and eventually vanish identically at $B \approx 2.3JS$.
At this high fields, the two-magnon density of states is zero at the energy of the (renormalized) edge mode, and the edge mode becomes a delta-peak.
Thus, the magnetic field has restored well-defined chiral edge excitations with infinite lifetime, and the LSWT bulk-boundary correspondence predicts the correct edge physics in the full spin model. Finally, we get back to the edge-edge hybridization discussed in Sec.~\ref{subsec:edge_edge_hybridization} and assess its behavior at finite magnetic fields.
From Fig.~\ref{fig:hybridization_vs_slab_size}(a), it is apparent that fields lower than $1.0JS$ do not have any significant impact on the edge-edge hybridization strength $v$.
In contrast, fields higher than $1.0JS$ lead to a much more rapid decay of $v$ with strip width $N_y$ than in the zero-field case.
The cause of this is the separation of the LSWT edge modes and the two-magnon continuum at $B \approx 1.0JS$, resulting in the suppression of the two-magnon mediated edge-edge coupling.
This result is consistent with the restoration of a well-defined topology in the high-field limit.

However, we remark that for typical exchange couplings of candidate materials (e.g., $J\approx 2\,\mathrm{meV}$ in CrI$_3$ \cite{chen_topological_2018}), the required magnetic field to reach the well-defined topology limit is of the order of several tens of Teslas.
An alternative strategy to find stable topological magnons is to look for strongly spin-anisotropic magnets with large anisotropy-induced spin-wave gaps, which can separate the single-magnon spectrum from the two-magnon continuum already in zero field. 
For the DMI ferromagnet at hand, examples for such an anisotropy would be the XXZ-type two-ion anisotropy or a single-ion easy-axis anisotropy.

\section{Discussion \& Conclusion}
\label{sec:relevance}

We have shown that magnon-magnon interactions can have destructive and non-universal effects on the chiral edge modes of the honeycomb ferromagnet with DMI.
Those range from strong decays to delocalization into the bulk to complete disappearance of the chiral edge modes from the spectrum.
Similar to Ref.~\onlinecite{chernyshev_damped_2016}, in our model the strength of the observed renormalization effects is tunable with the magnetization angle $\theta$.
At $\theta = 0$, the magnetization is aligned with the DMI vector ($D_\perp = 0$) and the resulting $U(1)$ symmetry gives rise to a magnon number conservation law~\footnote{Note that this relation between $U(1)$ symmetry and magnon number conservation only holds for ferromagnets.}.
As a result, spontaneous magnon decay is prohibited quantum-mechanically due to the absence of the corresponding transition matrix elements; the chiral edge modes are sharp and the bulk-boundary correspondence is exactly valid.
However, the $U(1)$ symmetric case is fine-tuned and expected to be violated in any real material.
In our model, the degree of violation is emulated by the size of $\theta$.
When gradually tilting the magnetization away from the DMI vector ($D_\perp \neq 0$), the bulk-boundary correspondence initially provides a fair approximation of the edge physics at small $\theta$.
However, for larger $\theta$ and thus larger $D_\perp$, the predictions of the bulk-boundary correspondence about the edge spectrum of the model become unreliable, and in some cases qualitatively wrong.
Crucially, estimating a critical angle up to which the LSWT bulk-boundary correspondence is still predictive is non-trivial due to non-universal decays, which depend not only on $\theta$ and the DMI magnitude $D$, but also on the two-magnon density of states, quantum matrix elements, edge terminations, and other model parameters such as external magnetic fields. Nevertheless, finding {\it soft} ferromagnets which allow a tuning of the direction of the magnetization, thus $\theta$, via a small external magnetic field would enable a tuning of interaction effects.

We expect that our findings for the honeycomb-lattice TMI have broad implications for topological magnons in other models which describe various real materials.
Examples are the kagome-lattice magnets Cu(1-3,bdc) \cite{chisnell_topological_2015, hirschberger_thermal_2015,chernyshev_damped_2016} and YMn$_6$Sn$_6$ \cite{zhang_topological_2020}, and the pyrochlore-lattice magnets, such as Lu$_2$V$_2$O$_7$ \cite{onose_observation_2010, zhang_topological_2013, Mook2016Weyl}.
Similar to the honeycomb model studied in this work, these systems utilize DMI to generate non-interacting topological magnons at the level of LSWT.
In analogy to the honeycomb model studied here, the DMI is capable of simultaneously introducing spin interactions of the form $S^{\pm} S^\parallel$, which violate magnon number conservation and enable spontaneous magnon decay \cite{chernyshev_damped_2016}.
Building upon our work, we propose that these non-conserving interactions pose a significant challenge to the stability of chiral edge modes in these models beyond LSWT.

We expect a similar breakdown of topological chiral edge magnons in field-polarized Kitaev magnets \cite{yongbaekkim_topological_2021}, such as $\alpha$-RuCl$_3$ \cite{czajka2023planar}. In the field-polarized limit, the Kitaev interaction introduces both topological magnon gaps and magnon-magnon interactions. In contrast to the DMI models discussed so far, the magnon number is already nonconserved at the LSWT level because of terms like $S^+ S^+$.
Therefore, a particle-hole-mixing Bogoliubov transform is necessary to diagonalize the LSWT.
Substituting the Bogoliubov eigenmodes into the interaction Hamiltonians $\hat H_{\ge 3}$ generates \emph{arbitrary} magnon-magnon vertices; for example, in addition to familiar three-magnon interactions $\sim a^\dagger a^\dagger a$ that we encountered in this work, four-magnon terms $\sim a^\dagger a^\dagger a^\dagger a$ also appear, allowing spontaneous decay into three final magnons \cite{Stephanovich_2011}.
Therefore, a rich renormalization of topological magnons is expected in magnetic fields too small to separate the single magnons from the two-magnon spectrum. Large enough magnetic fields stabilize topological chiral edge magnons, in accordance with Ref.~\onlinecite{mcclarty_topological_2018}.

So far, we considered collinear ferromagnetic phases, either stabilized in the ground state or by applying a strong magnetic field. These collinear phases have in common that the number-non-conserving interactions of the form $S^{\pm} S^\parallel$ necessarily originate from spin-orbit interactions (DMI, anisotropic exchange).
However, so do the topological magnon gaps at the LSWT level. Therefore, spin-orbit coupling is a blessing and a curse: it is necessary to open up a topological magnon gap in the first place, but at the same time installs magnon-magnon interactions that can lead to the demise of the topological magnon picture \cite{chernyshev_damped_2016}.

The effects of magnon-magnon interactions are even more pronounced in non-collinear magnets, in which topological magnons can arise due to a finite scalar spin chirality.
Examples are kagome antiferromagnets \cite{Owerre2017, Laurell2018, Mook2019coplanar,Lu2019anti}, pyrochlore antiferromagnets \cite{Li2016weyl, Laurell2017pyro}, and skyrmion crystals \cite{Hoogdalem2013, roldan2016topological, Diaz2019, Diaz2020, Mook2020skyrmion, weber2022topological}.
Due to the non-collinearity, the strength of the magnon-magnon interactions is given by the Heisenberg exchange and not just by spin-orbit coupling \cite{chernyshev_spin_2009, Chernyshev2015kagome}.
Consequently, a non-universal breakdown of the chiral edge modes should be expected unless their stability is explicitly demonstrated by a more detailed analysis.
Additionally, it should be noted that the high-field limit is not applicable in non-collinear magnets to restore well-defined chiral edge modes, since a sufficiently strong magnetic field will destroy the non-collinear order that is necessary for topological magnons to exist in the first place.

As a complement to our perturbation theory results, we propose a full quantum treatment of the honeycomb ferromagnet with DMI using numerical methods such as the density matrix renormalization group and time-evolution based on matrix product operators, which have already been used to study aspects of interacting magnon topology in Refs.~\onlinecite{mcclarty_topological_2018} and~\onlinecite{Gohlke2022}.
These methods can provide quantitative insight into the fate of the chiral edge modes even in the deep quantum limit $S=1/2$, where the perturbative truncation of the large-$S$ expansion after $\hat H_3$ is not formally justified. Importantly, non-perturbative simulations have already revealed mechanisms that help stabilize topological magnons: In the case of field reduction from large fields with separated one- and two-magnon spectra, the level repulsion between them prevents the single-particle modes from entering the continuum at the expense of the quasiparticle residue \cite{mcclarty_topological_2018, Gohlke2022,verresen_avoided_2019}.
The analysis of the breakdown mechanisms of chiral edge modes in the non-perturbative setting will be an important task for a more comprehensive understanding of magnon topology.

Furthermore, magnons are subject to a variety of interaction effects beyond the magnon-magnon interactions studied in this work.
An important interaction channel is the spin-lattice coupling \cite{Cong2022}.
While the resulting magnon-phonon coupling can itself lead to topological magnon-phonon hybrids in a non-interacting theory \cite{Takahashi2016, Zhang2019, Go2019, Sheikhi2021, luo2023evidence}, it has also been shown to significantly broaden the single-magnon modes in $\mathrm{CrGeTe_3}$ \cite{chen_anisotropic_2022} and has been suggested to lead to a substantial suppression of the magnon thermal Hall effect \cite{choi_sizable_nodate}.
Since these materials are thought to be described by variants of the honeycomb ferromagnet with DMI studied in this paper, the model provides a suitable platform for comparing the effects of magnon-phonon and magnon-magnon interactions on the chiral edge modes.
Furthermore, phonons are intrinsically gapless and cannot be gapped out by a magnetic field like magnons, which implies that magnon-phonon interactions could have potentially detrimental effects on the magnon topology even in the high-field limit, where magnon-magnon interactions are kinematically frozen out.
Therefore, we propose a detailed analysis of the stability of topological magnons in the presence of spin-lattice coupling for future studies.

We have derived our results for chiral edge states in two-dimensional TMI.
However, we expect the breakdown physics to carry over to other bosonic topological boundary modes, including helical magnon edge states in antiferromagnets \cite{Nakata2017, Mook2018dual, Kondo20192D}, magnon surface arcs of Weyl magnons \cite{Li2016weyl, Mook2016Weyl, Owerre2018weyl, Jian2018, Choi2019, Zhang2020weyl, Li2021weyl, Mook2021secondorder}, magnon drumhead states in magnon nodal line systems \cite{Mook2017nodal,Li2017nodal,Choi2019,Owerre2019,Hwang2020}, surface Dirac magnons \cite{Kondo20193D, Mook2021secondorder, Kondo2021}, and chiral hinge modes or corner states in second-order topological magnon systems \cite{Li2019soliton, Hirosawa2020, Li2020solitons, Mook2021secondorder, Park2021, Li2022higher, Karaki2023, Hua2023}.
The coupling to the two-magnon continuum is a universal effect caused by particle number non-conservation.
Consequently, such a coupling is not restricted to magnons.
We expect qualitatively identical effects for topological triplons \cite{Romhanyi2015,McClarty2017triplon, Anisimov2019, Bhowmick2021weyltriplon, Haldar2021}, topological phonons \cite{Peng2020, Li2021phonon, Wang2022, Xu2022}, and any other bosonic topological collective modes/quasiparticles without a number conservation law.
Therefore, our results provide further evidence that predictions of topological boundary modes obtained in a non-interacting theory must be carefully checked in the interacting case.
Since quantum interactions already occur at zero temperature, there is, in general, no low-temperature limit in which the non-interacting theory is trustworthy.

Finally, we comment on the implications of our results for spin and heat transport, in particular, for spin Nernst and thermal Hall effects.
The available expressions for the corresponding intrinsic transverse conductivities have been derived in the non-interacting limit, where the magnon Berry curvature fully determines the transverse transport \cite{katsura2010, matsumoto_rotational_2011, matsumoto_theoretical_2011, Cheng2016, Zyuzin2016}.
Recent experiments suggest that these expressions overestimate the thermal Hall effect of magnons and triplons \cite{Cairns2020, Suetsugu2022, choi_sizable_nodate}.
In principle, it is known how to express the conductivities in terms of spin operators \cite{Lee2015}, without the need for a Holstein--Primakoff transformation, and these expressions can be numerically evaluated in classical spin dynamics simulations \cite{Mook2016simulation, Mook2017simu2, Carnahan2021}.
However, to understand Hall-type transport in the deep quantum and low-temperature limit, a more comprehensive theory must be developed to provide insight into the microscopic many-body processes of the elementary excitations.
Since the three-particle interactions lead to a drastic renormalization of the topological features, we believe that they also provide the leading-order corrections to the Hall-type transport effects.

In conclusion, our results have important
implications for the design and implementation of TMI, as well as for a wide range of bosonic
many-body systems in the field of condensed matter physics. We discovered and quantified different breakdown mechanisms for chiral bosonic edge modes due to many-body interactions in conjunction with the absence of particle number conservation. On a positive note, we also discussed ways of restoring the stability which will hopefully guide future experiments for an unambiguous observation of chiral magnon edge modes.

\begin{acknowledgments}
We thank Pengcheng Dai, Paul~A. McClarty and Jeff~G. Rau for helpful discussions and especially the latter two for critical comments on the manuscript. J.~K. would like to thank Yanbai Zhang for contributions in the early stages of this project. J.~K. acknowledges support from the Imperial-TUM flagship partnership. The research is part of the Munich Quantum Valley, which is supported by the Bavarian state government with funds from the Hightech Agenda Bayern Plus.
This work was funded in part by the Deutsche Forschungsgemeinschaft (DFG, German
Research Foundation) - Project No.~504261060.
\end{acknowledgments}

\appendix

\section{Edge Terminations}
\label{sec:edge_terminations}

\fig{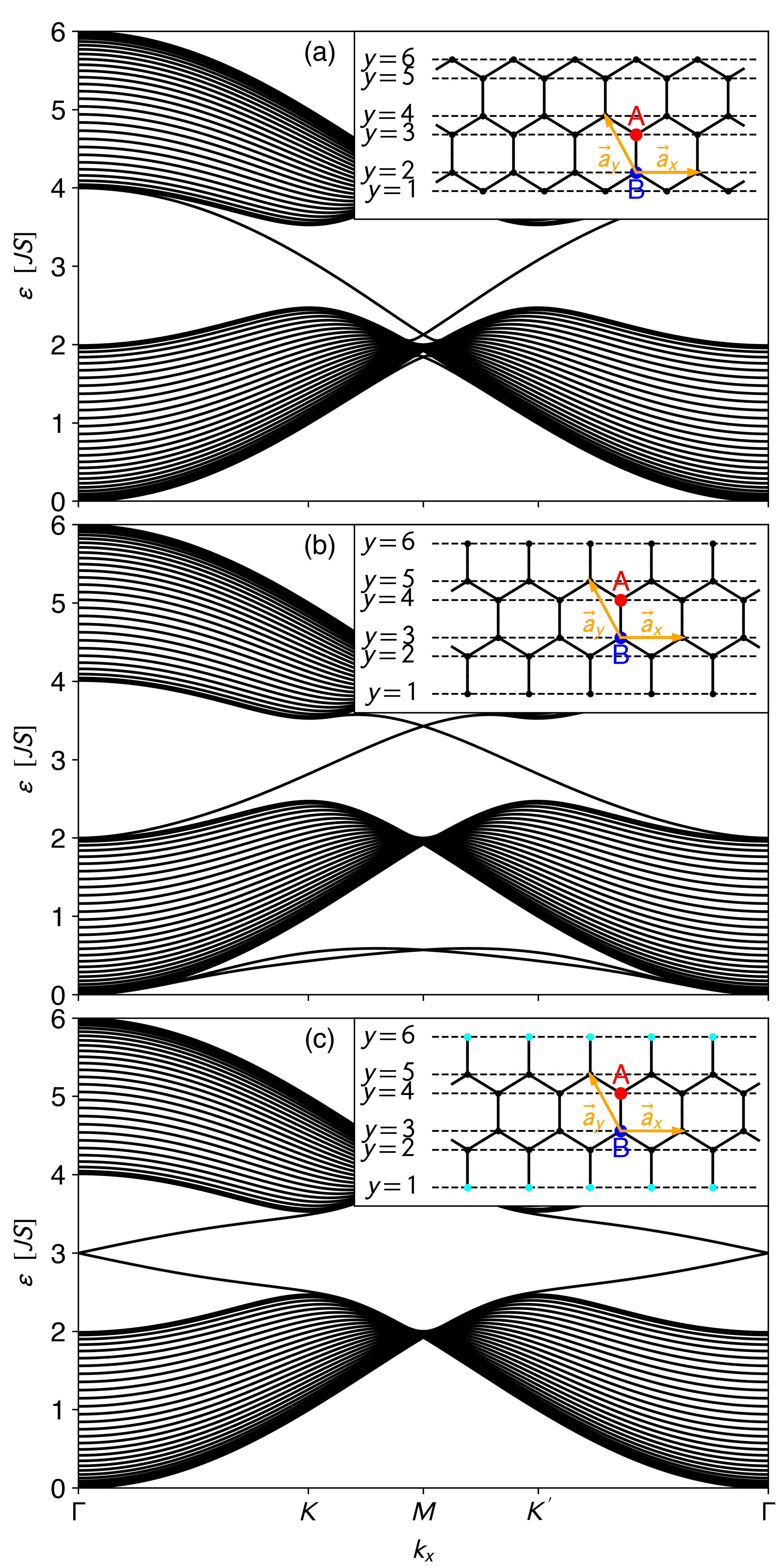}{LSWT band structures on the strip for (a) zig-zag edge termination, (b) dangling edge termination, and (c) dangling edge termination with edge chemical potential.
The edge modes in (a) and (b) lie in the vicinity of the $M$-point, while the edge modes in (c) are near the $\Gamma$-point.
In this study, we focus on the cases (a) and (c).
Also note the trivial low-energy edge modes in (b), which are pushed up into the lower bulk bands in (c) due to the edge chemical potential.
The insets show the corresponding lattices for $N_y = 3$.
Representative sites of the honeycomb sublattices $A$ and $B$ are indicated in red and blue, respectively.
The Bravais lattice vectors $\vec a_x$ and $\vec a_y$ are shown in orange, and the additional edge potential in (c) is marked by turquoise dots on the respective sites.
}

To elucidate the choice of edge terminations, we first label each site $i$ by its coordinates $(x,y)$ (see Fig.~\ref{fig:edge_terminations}).
To introduce open boundary conditions, we cut the honeycomb lattice along the $x$-direction at $y = 1$ and $y = 2N_y$.
The dangling edge termination is obtained when the lattice is cut in between the unit cells, while the zig-zag edge termination requires slicing the unit cells at the edge in half.

The number of nearest neighbors of a site $i$ at an edge (i.e., at $y=1, 2N_y$) is $z_i = 1$ for dangling, and $z_i = 2$ for zig-zag edges.
However, the LSWT edge mode spectra for the two edge terminations would be rather similar in that the edge modes all have momenta around the $M$-point [see Figs.~\ref{fig:edge_terminations}(a) and (b)].
Therefore, in the case of dangling edges, we add an ``edge chemical potential" of the form
\begin{align}
    2JS \sum_{x} \sum_{y\in\{1,2N_y\}} a_{x,y}^\dagger a_{x,y}
\end{align}
to the Hamiltonian (one could think of it as being generated by, e.g., an interfacial anisotropy).
This term effectively sets $z_i = 3$ at the edge, and shifts the dangling edge modes to the vicinity of the $\Gamma$-point [see Fig.~\ref{fig:edge_terminations}(c)].
The two sets of edge modes at different momenta help us demonstrate more clearly the non-universality of the edge mode decay (see Sec.~\ref{sec:renormalized_spectra}) and their delocalization into the bulk (see Sec.~\ref{sec:spontaneous_deloc_resonances}).

\section{Classical Spin Ground State on the Strip}
\label{app:classical_GS_strip}

The classical ground state of the investigated DMI-induced TMI on a torus geometry (i.e., only periodic boundary conditions) is known to be a completely polarized ferromagnet for weak enough DMI \cite{owerre_first_2016}.
However, this result does not trivially carry over to the strip geometry because the fewer number of nearest neighbors at the strip edges effectively reduce the ferromagnetic Heisenberg interaction felt by the edge spins.
For example, in the field-polarized Kitaev-Heisenberg model, spins at the edges tilt away from the ferromagnetic polarization axis of the bulk \cite{mcclarty_topological_2018}.
Therefore, one must carefully check that the DMI-induced TMI on the strip is \emph{not} subject to spin tilting and remains fully polarized at the edges.

To find the classical spin ground state on the strip, we need to find the spin configuration $\{\vec S_i\}$ which minimizes the classical Hamiltonian
\begin{align}
    H(\{\vec S_i\}) = \sum_{ij} \vec S_i^T \mathcal J_{ij} \vec S_j \nonumber
\end{align}
under the normalization constraint
\begin{align}
\label{eq:classical_GS_normalization}
    |\vec S_i|^2 = S^2 \quad \forall \text{ sites } i,
\end{align}
where $i,j$ are lattice site indices, $\mathcal J_{ij}$ is the $3N_xN_y\times 3N_xN_y$ spin interaction matrix between all sites $i$ and $j$, comprised of the nearest-neighbor Heisenberg coupling and the next-nearest-neighbor DMI, and $S$ is the local magnetic moment.
Note that the classical spins $\{\vec S_i\}$ and the Hamiltonian $H$ are not operator-valued but three-dimensional real vectors and a real scalar, respectively.

We first exploit the translational symmetry in $x$-direction through a Fourier transform $\vec S_y(k_x) = \sum_x e^{ik_x\sqrt 3 x} \vec S_{i=(x,y)}$, where the Fourier components fulfill $\vec S_y(k_x) = \vec S_y(-k_x)^*$.
The spin interaction matrices decompose into $3N_y \times 3N_y$-blocks $\mathcal J_{yy'}(k_x)$ so that the Hamiltonian becomes
\begin{align}
\label{eq:classical_Hamiltonian_momentum_space}
    H(\{\vec S_i\}) = \sumBZ{k_x} \sum_{yy'} \vec S_y^T(k_x) \mathcal J_{yy'}(k_x) \vec S_{y'}(k_x),
\end{align}
and the normalization constraint transforms to
\begin{align}
\label{eq:classical_GS_normalization_momentum_space}
    |\vec S_{i}|^2 = \sumBZ{k_x,q_x} e^{i(\vec k + \vec q)\sqrt 3 x} \vec S_y^T(k_x) \vec S_y(q_x) = S^2.
\end{align}
To satisfy Eq.~\eqref{eq:classical_GS_normalization_momentum_space} for all sites $i$, we need to enforce
\begin{gather}
\label{eq:classical_GS_normalization_momentum_space_unraveled1}
    \sumBZ{k_x} \vec S_y^T(-k_x) \vec S_y(k_x) = \sumBZ{k_x} |\vec S_y(k_x)|^2 = S^2, \\
\label{eq:classical_GS_normalization_momentum_space_unraveled2}
    \vec S_y^T(k_x) \vec S_y(q_x) = 0 \quad \text{for } k_x + q_x \neq 0
\end{gather}
for all $y$.
We employ two different numerical methods to carry out the constrained minimization.

\subsection{Luttinger-Tisza}

The idea is to temporarily ignore the normalization constraint \cite{luttinger_note_1951}.
Then, the problem reduces to the minimization of the quadratic form~\eqref{eq:classical_Hamiltonian_momentum_space}, which amounts to finding the lowest eigenvalue of each $k_x$-block.
Then, the ground state is the eigenvector corresponding to the overall lowest eigenvalue of all $k_x$-blocks.
However, this ground state is generally not normalized according to the constraint~\eqref{eq:classical_GS_normalization}, and therefore only an approximation.

As long as the DMI strength $D \lesssim 0.6J$ is weak enough, we find that $k_x=0$ yields the lowest overall eigenvalues, which are three-fold degenerate.
The corresponding eigenvectors have all spins pointing in the $x$-, $y$-, or $z$-direction, respectively.
This is characteristic for a ferromagnet whose polarization axis is completely arbitrary.
In particular, the spins do not tilt away from that axis near the edges.
This result is independent of the edge termination.

While the Luttinger-Tisza method yields qualitative evidence for the existence of a ferromagnet for finite DMI, the role of the normalization constraint remains unclear.

\subsection{Projected gradient descent}

This method provides a more rigorous treatment of the normalization constraint.
First, we convince ourselves that, even in the presence of the constraint, the ground state still is a configuration with fixed wavenumber $k_x$ (up to degeneracy).
Consider one spin configuration to be the superposition of different wavenumbers $\{K_x\}$, i.e., $\vec S_{i=(x,y)} = \sum_{K_x} e^{-i K_x \sqrt 3 x} \vec S_y(K_x)$, where $\vec S_y(K_x)$ is an eigenvector of $\mathcal J_{yy'}(K_x)$ with eigenvalue $\epsilon_{K_x}$.
Now define the wavenumber $Q_x$ so that $\epsilon_{Q_x}$ is the lowest of all eigenvalues $\{\epsilon_{K_x}\}$, and consider a second spin configuration $\vec S'_{i=(x,y)} = e^{-i Q_x \sqrt 3 x} \vec S'_y(Q_x)$ which has a unique wavenumber.
We show that the energy $H(\{\vec S_i\})$ of the first configuration is always greater than the energy $H(\{\vec S'_i\})$ of the second configuration:
\begin{align}
    H(\{\vec S_i\}) &= \sum_{K_x} \epsilon_{K_x} \sum_y |\vec S_{y}(K_x)|^2 \overset{\eqref{eq:classical_GS_normalization_momentum_space_unraveled1}}{>}
    \epsilon_{Q_x} \sum_y S^2, \nonumber \\
    &\overset{\eqref{eq:classical_GS_normalization_momentum_space_unraveled1}}{=} \epsilon_{Q_x} \sum_y |\vec S'_y(Q_x)|^2 = H(\{\vec S'_i\}). \nonumber
\end{align}
Therefore, a configuration with mixed wavenumbers can never be the ground state.
For a unique-wavenumber configuration such as $\{\vec S'_i\}$, the constraints~\eqref{eq:classical_GS_normalization_momentum_space_unraveled1} and \eqref{eq:classical_GS_normalization_momentum_space_unraveled2} simplify to
\begin{gather}
\label{eq:classical_GS_normalization_momentum_space_Qx1}
    |\vec S'_y(Q_x)|^2 = S^2, \\
\label{eq:classical_GS_normalization_momentum_space_Qx2}
    \vec {S'}_y^T(Q_x) \vec S'_y(Q_x) = 0 \quad \text{if } Q_x \neq 0
\end{gather}
for all $y$.

To compute $\vec S'_y(Q_x)$ for some $Q_x$ using the projected gradient descent method, we start from a random initial spin configuration and improve it iteratively until convergence is achieved up to a tolerance.
Each iteration, we evaluate the gradient $\delta H / \delta \vec S_y(Q_x)$ at the previous configuration, perform a gradient descent step with step size $0.3$, and project the thus obtained spin configuration onto the manifold defined by the constraints~\eqref{eq:classical_GS_normalization_momentum_space_Qx1} and \eqref{eq:classical_GS_normalization_momentum_space_Qx2}.
Finally, we compare the obtained energies in each $Q_x$-block to find the spin configuration with the overall lowest energy (i.e., the ground state spin configuration).

We obtain that, as long as $D \lesssim 0.7J$, the converged configuration for $Q_x = 0$ has the lowest energy.
It describes the same ferromagnetic alignment as the Luttinger-Tisza result.
Hence, we conclude that the strip ground state of our model is indeed fully polarized and does not exhibit spin tilting at the edges in this parameter regime.

\section{Magnon-Magnon Vertex Factors}
\label{app:magnon_magnon_vertex}
To illustrate the derivation of the vertex factors in Eq.~\eqref{eq:cubic_hamiltonian_momentum_space}, we focus on the dangling edge termination; the equations can be swiftly generalized to the zig-zag edge termination.

The convention for the coordinates of a lattice site $i = (x,y)$ is as in App.~\ref{sec:edge_terminations}.
We introduce the set $\mathrm{nnn}(x,y)$, which contains the distance vectors to the next-nearest neighbor sites of $i$, expressed in terms of the honeycomb Bravais lattice vectors.
That is,
\begin{align}
	\mathrm{nnn}(x,y\ \mathrm{even}) &= \begin{cases}
		\{(-1, 0), (1, 1), (0, -1)\} & 2 \le y/2 \le N_y-1 \\
		\{(-1, 0), (1, 1)\} & y/2 = 1 \\
		\{(-1, 0), (0, -1)\} & y/2 = N_y
	\end{cases} \nonumber \\
	\mathrm{nnn}(x,y\ \mathrm{odd}) &= \{-\vec v_\mathrm{nnn}\ |\ \vec v_\mathrm{nnn} \in \mathrm{nnn}(x,y+1)\} \nonumber 
\end{align}
Note the case distinction between even and odd $y$, which correspond to sites on the $A$ and $B$ sublattices, respectively.
Due to translational invariance in $x$-direction, $\mathrm{nnn}(x,y)$ does not depend on $x$, so we write $\mathrm{nnn}(y)$ from now on.
With these definitions, the sum over all next-nearest neighbors $\sum_{\nnn{ij}}$ becomes $\sum_{x,y} \sum_{\substack{(\delta x,\delta y) \in \mathrm{nnn}(y)}}$.

To compute the vertex factors, we start from Eq.~\eqref{eq:cubic_hamiltonian_real_space} and plug in the LSWT diagonalization transform \eqref{eq:LSWT_diagonalization}.
We subsequently use the completeness relation $\sum_x e^{i(p_x+q_x-k_x)\sqrt 3 x} = N_x \,\delta_{p_x+q_x-k_x,0}$:
\newcommand{\bidx}[1]{
	\IfEqCase{#1}{
		{3}{n}
		{4}{m}
		{5}{l}
	}
}	
\begin{align}
	\hat H_3 &= \frac{D_\perp \sqrt{S}}{\sqrt 2} \sum_{\nnn{ij}} \br{i a_i^\dagger a_j^\dagger \br{a_j - a_i} + \hc} \nonumber \\
	&= \frac{D_\perp \sqrt{S}}{\sqrt 2} \sum_{x,y} \sum_{\substack{(\delta x,\delta y) \in \\ \mathrm{nnn}(y)}} \br{i a_{x,y}^\dagger a_{x+\delta x,y+\delta y}^\dagger \br{a_{x+\delta x,y+\delta y} - a_{x,y}} + \hc} \nonumber \\
    &= \frac{D_\perp \sqrt S}{\sqrt{N_x}} \sum_{\bidx 3\bidx 4\bidx 5} \sumBZ{k_x, q_x} f_{k_x,q_x,k_x-q_x}^{\bidx 3\bidx 4\bidx 5} a_{k_x-q_x,\bidx 3}^\dagger a_{q_x,\bidx 4}^\dagger a_{k_x,\bidx 5}. \nonumber
\end{align}
$f_{k_x,q_x,k_x-q_x}^{\bidx 3\bidx 4\bidx 5}$ is the precursor of our final vertex factor $g_{k_xq_x}^{\bidx 3\bidx 4\bidx 5}$:
\begin{align}
	f_{k_x,q_x,k_x-q_x}^{\bidx 3\bidx 4\bidx 5} &= \frac{i}{\sqrt 2} \sum_{y} \sum_{\substack{(\delta x,\delta y) \in \\ \mathrm{nnn}(y)}} \mathcal U_{\bidx 3 y}^*(k_x-q_x) \mathcal U^*_{\bidx 4,y+\delta}(q_x) \nonumber\\
    &\cdot \br{\mathcal U_{\bidx 5,y+\delta y}(k_x) e^{-ik_x\sqrt 3\delta x} - \mathcal U_{\bidx 5 y}(k_x) e^{-i(k_x-q_x)\sqrt 3 \delta x}}. \nonumber
\end{align}
It is however not yet symmetric under exchanging $(q_x, \bidx 4) \leftrightarrow (k_x-q_x, \bidx 3)$.
Therefore, we symmetrize $\hat H_3$:
\begin{footnotesize}
\begin{align}
	\hat H_3 &= \frac{D_\perp \sqrt S}{\sqrt{N_x}} \sqrt S \sum_{\bidx 3\bidx 4\bidx 5} \sumBZ{k_x, q_x} f_{k_x,q_x,k_x-q_x}^{\bidx 3\bidx 4\bidx 5} a_{k_x-q_x,\bidx 3}^\dagger a_{q_x,\bidx 4}^\dagger a_{k_x,\bidx 5} \nonumber \\
	&= \frac{D_\perp \sqrt S}{\sqrt{N_x}} \sum_{\bidx 3\bidx 4\bidx 5} \sumBZ{k_x, q_x} \underbrace{\frac{1}{2} \br{f_{k_x,q_x,k_x-q_x}^{\bidx 3\bidx 4\bidx 5} + f_{k_x,k_x-q_x,q_x}^{\bidx 4\bidx 3\bidx 5}}}_{g_{k_x q_x}^{\bidx 3\bidx 4\bidx 5}} a_{k_x-q_x,\bidx 3}^\dagger a_{q_x,\bidx 4}^\dagger a_{k_x,\bidx 5}, \nonumber
\end{align}
\end{footnotesize}%
where we swapped $\bidx 4 \leftrightarrow \bidx 3$ and shifted $q_x \rightarrow k_x-q_x$ in the second summand (if $k_x-q_x$ exceeds the Brillouin zone, we fold it back by adding or subtracting $2\pi/\sqrt 3$, which is the size of the 1D Brillouin zone of the strip).
Finally, we arrive at
\begin{align}
\label{eq:cubic_vertex_factors}
	g_{k_x q_x}^{\bidx 3\bidx 4\bidx 5} &= \frac{i}{2\sqrt 2} \sum_{y=1}^{2N_y} \sum_{\substack{(\delta x,\delta y) \in \mathrm{nnn}(y)}} \\
    \bigg(
	&+ e^{-i(k_x-q_x)\sqrt 3\delta x} \mathcal U_{\bidx 3 y}^*(k_x-q_x) \mathcal U_{\bidx 4,y+\delta y}^*(q_x) \mathcal U_{\bidx 5,y+\delta y,s}(k_x) \nonumber \\
	&- e^{iq_x\sqrt 3\delta x} \mathcal U_{\bidx 3 y}^*(k_x-q_x) \mathcal U_{\bidx 4,y+\delta y}^*(q_x) \mathcal U_{\bidx 5 y}(k_x)
	\nonumber \\
	\nonumber
	&+ e^{-iq_x\sqrt 3\delta x} \mathcal U_{\bidx 3,y+\delta y}^*(k_x-q_x) \mathcal U_{\bidx 4 y}^*(q_x) \mathcal U_{\bidx 5,y+\delta y}(k_x) \\
	&- e^{i(k_x-q_x)\sqrt 3\delta x} \mathcal U_{\bidx 3,y+\delta y}^*(k_x-q_x) \mathcal U_{\bidx 4 y}^*(q_x) \mathcal U_{\bidx 5 y}(k_x) \bigg). \nonumber
\end{align}

\section{Validity of the Perturbative Treatment}
\label{app:validity_pert_th}
We deem our perturbative approach valid if the self-energy scale is much smaller than the scale of the LSWT energies $\epsilon \sim JS$.
We read off the self-energy scale from Eq.~\eqref{eq:self_energy_one_loop}:
\begin{align}
    \hat\Sigma \sim 2D_\perp^2 S \sum \frac{|g|^2}{\omega - \epsilon - \epsilon} \sim \frac{D_\perp^2 S}{JS} = \frac{D_\perp^2}{J}. \nonumber
\end{align}
Therefore, perturbation theory is naively valid if
\begin{align}
    \frac{D_\perp^2}{J^2 S} \ll 1. \nonumber
\end{align}
For the largest interaction strength $D_\perp = 0.4J\sqrt S$ studied in this work, we have $D_\perp^2/(J^2 S) = 0.16$, which is on the verge of being much less than one.
However, any two-loop correction to the self-energy would only enter at order $D_\perp^4/(J^4 S^2) = 0.0256$, which is small enough to reasonably neglect.

Furthermore, we need to check that $D_\perp$ is small enough to ensure a ferromagnetic ground state.
In Sec.~\ref{sec:model}, we found that the model is a ferromagnet if
\begin{align}
    D = \sqrt{D_\parallel^2 + D_\perp^2} \lesssim 0.7J. \nonumber
\end{align}
Plugging in the value $D_\parallel = 0.1 JS$ which we used throughout this work, we obtain
\begin{align}
    D_\perp^2 \lesssim 0.48 J^2. \nonumber
\end{align}
Using the maximal investigated interaction strength $D_\perp = 0.4J\sqrt S$, we find that the ground state is ferromagnetic for $S \lesssim 3$.
This condition is fulfilled for many candidate materials of the honeycomb ferromagnet with DMI like CrI$_3$, where the magnetic Cr ion has spin $S=3/2$ \cite{chen_topological_2018}.

We remark that this analysis completely neglects the role of the multi-magnon density of states, which implicitly enter in the self-energy expression at all orders via the corresponding multi-magnon propagators.
For example, singularities in the multi-magnon densities of states could render the one-loop approximation quantitatively inaccurate.
However, we expect that the non-universal nature of the breakdown mechanisms of the edge mode will qualitatively remain the same.

\section{On-Shell Approximation of the Decay Rates}
\label{app:decay_rate_os_approx}
The general idea of the on-shell approximation is to obtain a first-order approximation to the self-consistent solution of the pole condition \eqref{eq:pole_equation} by exploiting the fact that the self-energy is small compared to the LSWT energies in the context of the large-$S$ expansion (``$\SE/\hat\epsilon \sim 1/S$").

Firstly, we demonstrate that, at first order in $1/S$, we can usually neglect the off-diagonal self-energy terms.
We split up the self-energy into a diagonal and an off-diagonal part, i.e. $\SE \equiv \SE_\mathrm{diag} + \SE_\mathrm{offdiag}$.
Next, for convenience, we define a short-hand for the diagonal part of the interacting Green function:
\begin{align}
    \GF_\mathrm{diag}(\omega, k_x) \equiv \left((\omega + i0^+)\hat{\mathbbm 1} - \hat\epsilon_{k_x} - \SE_\mathrm{diag}(\omega^*, k_x)\right)^{-1}. \nonumber
\end{align}
With this, we Taylor-expand the pole condition up to order $1/S$:
\begin{align}
    0 &= \det\left((\tilde\epsilon_{k_x,n} + i0^+) \hat{\mathbbm 1} - \SE(\tilde\epsilon_{k_x,n}^*, k_x)\right) \nonumber \\
    &= \det\left(\GF_\mathrm{diag}^{-1}(\tilde\epsilon_{k_x,n}, k_x) - \SE_\mathrm{offdiag}(\tilde\epsilon_{k_x,n}^*, k_x)\right) \nonumber \\
    &= \det\left(\GF_\mathrm{diag}^{-1}\right) \left(1 - \tr\left(\GF_\mathrm{diag} \SE_\mathrm{offdiag}\right) + \mathcal{O}\left(1/S^2\right)\right). \nonumber
\end{align}
Since the product of the diagonal matrix $\GF_\mathrm{diag}$ and the purely off-diagonal matrix $\SE_\mathrm{offdiag}$ is traceless, the pole equation simplifies to $\det\left(\GF_\mathrm{diag}(\tilde\epsilon_{k_x,n}, k_x)\right) = 0$.
Dividing away the non-singular factors of the determinant, we obtain
\begin{align}
\label{eq:pole_eq_diagonal_self_energy}
    0 = \tilde\epsilon_{k_x,n} - \epsilon_{k_x,n} - \SE_{nn}(\tilde\epsilon_{k_x,n}^*, k_x) + \mathcal O\left(1/S\right).
\end{align}
This last step is only possible if we assume that the renormalized energies are non-degenerate.
If they are however degenerate, we would need to take into account the off-diagonal self-energy terms at higher order in $1/S$, which would remove the degeneracy by hybridization and level-splitting.
For the sake of simplicity, we neglect this effect for the on-shell decay rates.
However, when we later address the hybridization of edge modes on opposite sides of the system through an effective non-Hermitian on-shell Hamiltonian, we must keep the relevant off-diagonal terms (see Sec.~\ref{subsec:edge_edge_hybridization}).

Secondly, we note that the Eq.~\eqref{eq:pole_eq_diagonal_self_energy} implies that the renormalized energies $\tilde\epsilon_{k_x,n}$ only differ from the LSWT energies $\epsilon_{k_x,n}$ by a term of subleading order in $1/S$, i.e., $(\tilde\epsilon_{k_x,n} - \epsilon_{k_x,n})/\epsilon_{k_x,n} \sim 1/S$.
Therefore, we can already obtain $\tilde\epsilon_{k_x,n}$ up to order $1/S$ by doing one iteration step:
\begin{align}
    0 &= \tilde\epsilon_{k_x,n} - \epsilon_{k_x,n} - \SE_{nn}(\epsilon_{k_x,n} + \SE^*_{nn}(\tilde\epsilon_{k_x,n}, k_x), k_x) + \mathcal O\left(1/S\right) \nonumber \\
    &= \tilde\epsilon_{k_x,n} - \epsilon_{k_x,n} - \SE_{nn}(\epsilon_{k_x,n}, k_x) + \mathcal O\left(1/S\right). \nonumber 
\end{align}
This yields the familiar result for the on-shell decay rate $\Gamma^\mathrm{os}_{k_x,n}$:
\begin{align}
    \Gamma_{k_x,n} = -2\im \tilde\epsilon_{k_x,n} = \underbrace{-2\im \SE_{nn}(\epsilon_{k_x,n}, k_x)}_{\Gamma^\mathrm{os}_{k_x,n}} + \mathcal O\left(1/S\right). \nonumber 
\end{align}

\section{Perturbative Analysis of the $y$-Resolved Spectral Function}
\label{app:deloc_param_perturbative}

\newcommand{\treelevelvertex}[3]{\begin{tikzpicture}[baseline=(current bounding box.center)]
	\filldraw[black] (-0.1,0) circle (0pt) node[anchor=south]{\textit{#1}};
	\draw (-0.2,0) -- (0.5,0);
	\filldraw[black] (0.5,0) circle (2pt) node[anchor=west]{};
	\draw (0.5,0) -- (1, 0.5);
	\filldraw[black] (1, 0.5) circle (0pt) node[anchor=west]{\textit{#2}};
	\draw (0.5,0) -- (1, -0.5);
	\filldraw[black] (1, -0.5) circle (0pt) node[anchor=west]{\textit{#3}};
\end{tikzpicture}}

\newcommand{\treelevelvertexinv}[3]{\begin{tikzpicture}[baseline=(current bounding box.center)]
	\filldraw[black] (0, 0.5) circle (0pt) node[anchor=east]{\textit{#1}};
	\draw (0,0.5) -- (0.5, 0);
	\filldraw[black] (0, -0.5) circle (0pt) node[anchor=east]{\textit{#2}};
	\draw (0,-0.5) -- (0.5, 0);
	\filldraw[black] (0.5,0) circle (2pt) node[anchor=west]{};
	\draw (0.5,0) -- (1.2,0);
	\filldraw[black] (1.1,0) circle (0pt) node[anchor=south]{\textit{#3}};
\end{tikzpicture}}

\newcommand{\oneloopselfenergy}[4]{\begin{tikzpicture}[baseline=(current bounding box.center)]
	\filldraw[black] (0.1,0) circle (0pt) node[anchor=south]{\textit{#1}};
	\draw (0,0) -- (0.5,0);
	\filldraw[black] (0.5,0) circle (2pt) node[anchor=west]{};
	\draw (0.5,0) arc [
		start angle=-180,
		end angle=180,
		x radius=0.5,
		y radius=0.3
	];
	\filldraw[black] (1,0.3) circle (0pt) node[anchor=south]{\textit{#2}};
	\filldraw[black] (1,-0.3) circle (0pt) node[anchor=north]{\textit{#3}};
	\filldraw[black] (1.5,0) circle (2pt) node[anchor=west]{};
	\draw (1.5,0) -- (2,0);
	\filldraw[black] (1.9,0) circle (0pt) node[anchor=south]{\textit{#4}};
\end{tikzpicture}}

To understand the origins of the delocalized spectral response at the edge mode energies, we investigate which states contribute to the propagation of an edge magnon.
We first derive the exact Lehmann representation of the $y$-resolved spectral function based on the Lehmann representation of the renormalized zero-temperature single-magnon Green function:
\begin{align}
    \GF_{nm}(\omega, k_x) &= \sum_\phi \frac{\braket{0 | a_{k_x,m} | \phi} \braket{\phi | a_{k_x,n}^\dagger | 0}}{\omega - E_\phi + i0^+}, \nonumber
\end{align}
where $n$ and $m$ are LSWT band indices, and $\phi$ labels a complete set of renormalized eigenstates of the full Hamiltonian with energies $E_\phi$.
Using this, we find
\begin{footnotesize}
    \begin{align}
        \mathcal A_y(\omega, k_x) &= -\frac{1}{\pi} \im\left(\sum_{nm} \mathcal U_{yn}^\dagger(k_x) \sum_\phi \frac{\braket{0 | a_{k_x,m} | \phi} \braket{\phi | a_{k_x,n}^\dagger | 0}}{\omega - E_\phi + i0^+} \mathcal U_{my}(k_x) \right) \nonumber \\
        &= -\frac{1}{\pi} \im \sum_\phi \frac{\left|\braket{0 | a_{k_x,y} | \phi}\right|^2}{\omega - E_\phi + i0^+} \nonumber \\
        \label{eq:y_resolved_spectral_function_Lehmann_rep}
        &= \sum_\phi \left|\braket{0 | a_{k_x,y} | \phi}\right|^2 \delta(\omega - E_\phi),
    \end{align}
\end{footnotesize}%
where the $\delta$-function should be interpreted as a Lorentzian in case $E_\phi$ is complex.
Restricting ourselves to the one- and two-magnon sectors, $\ket{\phi}$ can be either a single-magnon state or a two-magnon state.
We express $\ket\phi$ using standard perturbation theory.
If $\ket\phi$ is a renormalized single-magnon state (momentum $k_x$, band index $n$):
\begin{align}
    \ket\phi &= \sqrt{Z_{k_x,n}}\, a_{k_x,n}^\dagger \ket 0 + \sum_{q_x,ab} C_{1, k_x q_x}^{nab} a_{q_x,a}^\dagger a_{k_x-q_x,b}^\dagger \ket 0 \nonumber \\
    &\ + \sum_{m \neq n} C_{2, k_x}^{nm} a_{k_x,m}^\dagger \ket 0  + \mathcal O(\hat{H}_3^3), \nonumber 
\end{align}
where $Z_{k_x,n}$ is the quasiparticle weight of the state.
The expansion coefficients read
\begin{footnotesize}
    \begin{align}
        \hspace{2em}
        C_{1, k_xq_x}^{nab} &= \frac{\braket{0 | a_{q_x,a} a_{k_x-q_x,b} \hat{H}_3 a_{k_x,n}^\dagger | 0 }}{\epsilon_{k_x,n} - \epsilon_{q_x,a} - \epsilon_{k_x-q_x,b} + i0^+} \nonumber \\
        &= \frac{1}{\epsilon_{k_x,n} - \epsilon_{q_x,a} - \epsilon_{k_x-q_x,b} + i0^+} \times \treelevelvertex{n}{a}{b} \nonumber \\[1.0\baselineskip]
        C_{2, k_x}^{nm} &= \sum_{q_x,ab} \frac{\braket{0 | a_{k_x,m} \hat{H}_3 a_{q_x,a}^\dagger a_{k_x-q_x,b}^\dagger | 0 } \braket{0 | a_{q_x,a} a_{k_x-q_x,b} \hat{H}_3 a_{k_x,n}^\dagger | 0 }}{(\epsilon_{k_x,n} - \epsilon_{k_x,m})(\epsilon_{k_x,n} - \epsilon_{q_x,a} - \epsilon_{k_x-q_x,b} + i0^+)} \nonumber \\
        &= \frac{1}{\epsilon_{k_x,n} - \epsilon_{k_x,m}} \times \oneloopselfenergy{n}{a}{b}{m}. \nonumber
    \end{align}
\end{footnotesize}%
If $\ket\phi$ is a renormalized two-magnon state (total momentum $k_x$):
\begin{align}
    \ket\phi &= a_{q_x,a}^\dagger a_{k_x-q_x,b}^\dagger \ket 0  + \sum_m D_{k_xq_x}^{mab} a_{k_x,m}^\dagger \ket 0  + \mathcal O(\hat{H}_3^2), \nonumber 
\end{align}
where
\begin{footnotesize}
    \begin{align}
        D_{k_xq_x}^{mab} &= \frac{\braket{0 | a_{k_x,m} \hat{H}_3 a_{q_x,a}^\dagger a_{k_x-q_x,b}^\dagger | 0}}{\epsilon_{q_x,a} + \epsilon_{k_x-q_x,b} - \epsilon_{k_x,m} + i0^+} \nonumber \\
        &= \frac{1}{\epsilon_{q_x,a} + \epsilon_{k_x-q_x,b} - \epsilon_{k_x,m} + i0^+} \times \treelevelvertexinv{a}{b}{m}. \nonumber
    \end{align}
\end{footnotesize}%

Plugging these expressions for $\ket\phi$ into Eq.~\eqref{eq:y_resolved_spectral_function_Lehmann_rep} and restricting ourselves to frequencies $\omega$ inside the band gap near the renormalized energy of the edge mode at the lower edge $\omega \approx E_{k_x,l}$, we find
\begin{footnotesize}
    \begin{gather}
        \mathcal A_y(\omega \approx E_{k_x,l}, k_x) \approx \left| \mathcal U_{ly} \sqrt{Z_{k_x,l}} + \sum_{m\neq l} \mathcal U_{my} C_{2,k_x}^{lm} \right|^2 \delta(\omega - E_{k_x,l}) \nonumber \\
        \label{eq:y_resolved_spectral_function_Lehmann_rep_perturbative}
        + \sum_{\substack{q_x,ab \\ \in \text{band gap}}} \left| \sum_m \mathcal U_{my} D_{k_xq_x}^{mab} \right|^2 \delta(\omega - \epsilon_{q_x,a} + \epsilon_{k_x-q_x,b}),
    \end{gather}
\end{footnotesize}%
where $q_x,ab \in \text{band gap}$ means that the energy $\epsilon_{q_x,a} + \epsilon_{k_x-q_x,b}$ of the corresponding two-magnon state should be inside the band gap.
The first summand in Eq.~\eqref{eq:y_resolved_spectral_function_Lehmann_rep_perturbative} is the coherent response of the edge mode resonance.
It is composed of the original LSWT edge mode (wavefunction $\mathcal U_{ly}$) damped by the wavefunction renormalization constant, and admixtures of single-magnon bulk states (wavefunctions $\mathcal U_{my}$).
The second summand corresponds to the incoherent two-magnon background due to the hybridization of single-magnon edge and two-magnon bulk states.
Both contributions generate the spectral response inside the bulk shown in Fig.~\ref{fig:spectra_real_space}.

Comparing the interacting $y$-resolved spectral function in Eq.~\eqref{eq:y_resolved_spectral_function_Lehmann_rep_perturbative} to the non-interacting one
\begin{align}
    \mathcal A^0_y(\omega \approx E_{k_x,l}, k_x) &= \left| \mathcal U_{ly} \right|^2 \delta(\omega - E_{k_x,l}), \nonumber
\end{align}
we see that the first non-vanishing correction to the coherent (incoherent) part is proportional to $C_{2,k_x}^{lm}$ ($D_{k_xq_x}^{mab}$), both of which are of order $\mathcal O(\hat{H}_3^2) \sim \mathcal O(D_\perp^2)$.

To single out the hybridization of the edge state (band index $l$) with single-magnon bulk states (band indices $m$), we can explicitly compute their overlaps $C_{2,k_x}^{lm}$ up to order $\mathcal O(D_\perp^2)$.
Then, the quantity
\begin{align}
    \sum_{m\in\text{bulk}} \left| C_{2,k_x}^{lm} \right|^2 = \sum_{m\in\text{bulk}} \left| \frac{\hat\Sigma_{lm}(\epsilon_{k_x,l}, k_x)}{\epsilon_{k_x,l} - \epsilon_{k_x,m}} \right|^2
\end{align}
describes the percentage of single-magnon bulk states in the renormalized edge mode wavefunction.
Unsurprisingly, it depends on the off-diagonal edge-bulk-coupling components of the self-energy matrix.

\bibliography{paper_v1}

\end{document}